# Rank-reduced coupled-cluster III. Tensor hypercontraction of the doubles amplitudes

Edward G. Hohenstein,[1,2*] B. Scott Fales,[1,2] Robert M. Parrish[3] and Todd J. Martínez[1,2]

[1]Department of Chemistry and The PULSE Institute, Stanford University, Stanford, CA 94305

[2]SLAC National Accelerator Laboratory, 2575 Sand Hill Road, Menlo Park, CA 94025

[3]QC Ware Corporation, Palo Alto, CA, 94303

*egh4@slac.stanford.edu

**Abstract:**

We develop a quartic-scaling implementation of coupled-cluster singles and doubles based on low-rank tensor hypercontraction (THC) factorizations of both the electron repulsion integrals (ERIs) and the doubles amplitudes. This extends our rank-reduced coupled-cluster method to incorporate higher-order tensor factorizations. The THC factorization of the doubles amplitudes accounts for most of the gain in computational efficiency as it is sufficient, in conjunction with a Cholesky decomposition of the ERIs, to reduce the computational complexity of most contributions to the CCSD amplitude equations. Further THC factorization of the ERIs reduces the complexity of certain terms arising from nested commutators between the doubles excitation operator and the two-electron operator. We implement this new algorithm using graphical processing units (GPUs) and demonstrate that it enables CCSD calculations for molecules with 250 atoms and 2500 basis functions using a single computer node. Further, we show that the new method computes correlation energies with comparable accuracy to the underlying RR-CCSD method.

## I. Introduction

Coupled-cluster (CC) theory remains perhaps the most successful single-reference method for describing electron correlation in molecules.[1] Much of this success is due to the rigorous size extensivity and size consistency of the CC energy that results from the exponential wavefunction ansatz.[2-4] This scaling behavior transfers to molecular properties derived from coupled-cluster wavefunctions by following response formalisms.[5] The principal disadvantage of CC-based methods is the steep scaling of the computational cost with system size. The coupled-cluster singles and doubles (CCSD) method[6-8] scales with the sixth-power of system size, $O(N^6)$, and the perturbative triples correction, (T),[9] which is required for high accuracy predictions, scales with the seventh-power of system size, $O(N^7)$. Further, the storage of the CC doubles amplitudes grows with the fourth-power of system size. In practice, this steep asymptotic scaling limits the applicability of canonical CCSD computations to molecules with up to roughly 100 atoms and 1000 basis functions using commonly available parallel computing resources.[10-16]

Many approaches have been developed to combat the high computational complexity of CC methods by exploiting sparsity in the CC wavefunction and the operators that underlie all of electronic structure theory. Broadly speaking, there are two types of sparsity that can be considered: elementwise sparsity and rank sparsity. Elementwise sparsity is typically a manifestation of spatial locality. In a Gaussian atomic orbital basis set, elementwise sparsity appears in integrals over the usual one- and two-electron operators with pairs of orbitals whose centers are spatially distant. For electron repulsion integrals (ERIs), the Cauchy-Schwarz inequality provides a bound for the magnitude of an ERI,[17] thus allowing the elementwise sparsity to be exploited in integral direct algorithms.[18] Alternatively, the rank-sparsity of the ERIs can be considered. Such methods seek to find hidden factorizable structures that expose low numerical rank in the tensors that appear in electronic structure theory. For example, the density-fitting (DF)[17, 19-22] and Cholesky decomposition (CD)[23-25] approximations of the ERIs construct low-rank representations of these four-index tensors as products of three-index tensors. This factorized structure has proven extremely useful in accelerating many-body perturbation theory (MBPT) computations. Further factorization of the ERIs is possible following the pseudospectral approximation (PS)[26-30] and the related chain-of-spheres method (COS).[31-33] The most flexible low-rank factorization of the ERIs is provided by tensor hypercontraction (THC), where the four-index ERI tensor is decomposed into a product of five matrices.[34-36] Unfortunately, exploiting rank and/or elementwise sparsity in the ERIs alone does not appear to reduce the computational complexity of CCSD.

Elementwise sparsity is not immediately exposed in the CC wavefunction amplitudes, because CC methods are usually formulated in a canonical delocalized molecular orbital basis. It is possible to formulate CC methods in an atomic orbital (AO) basis where elementwise sparsity is apparent.[37-38] However, efforts to develop such AO-based methods were largely stymied by the length scales required before such a representation provides a practical advantage. An alternative approach is to work in a basis of local molecular orbitals.[39-42] This allows elementwise sparsity to be introduced in two key areas. By localizing the orbitals, neglible contributions from distant electron pairs can be neglected (screened) and small, but non-negligible, contributions can be treated approximately.[43] Additionally, the number of virtual orbitals required to correlate an electron pair can be reduced by projecting into a reduced atomic orbital basis – these virtual orbitals are known as projected atomic orbitals (PAOs).[39] Schütz and Werner pioneered the application of these local correlation methods to CC theory and this strategy has had more practical success than pure AO-based methods.[44-47] The disadvantage of these methods is the appearance of discontinuities on the global potential energy surface resulting from the numerical thresholds introduced to screen negligible contributions.[48] Lowering these thresholds can remove numerical artifacts, but also increases the computational expense. Additionally, a clear path to evaluate molecular response properties with reduced-scaling local CC methods, while still preserving their computational advantage, has not been established.[48-52] Finally, we note that methods such as the cluster-in-molecule local correlation approach[53-55] as well as other fragmentation-based schemes[56-58] can also be viewed as attempts to leverage elementwise sparsity.

More recently, CC methods based on pair natural orbitals have emerged, driven in large part by the work of Neese and Werner.[59-66] These methods leverage a combination of elementwise and rank sparsity. An orbital localization is performed so that approximations can be applied to electron pairs based on their spatial separation; this enables elementwise sparsity in the doubles amplitudes to be leveraged.[67] The virtual orbitals for each correlated electron pair are the pair natural orbitals (PNOs)[68-71] obtained from the eigenvectors of the pair density (often using a second-order Moller-Plesset perturbation theory (MP2) approximation). By truncating the pair natural orbital space on the basis of their occupancy, a low-rank approximation of the doubles amplitudes for each electron pair is introduced. Further, the introduction of orbital domains – i.e. domain based local pair natural orbitals (DLPNOs)[72] – allows spatial sparsity to be fully exploited while retaining the low-rank representation of the doubles amplitudes. The introduction of PNOs should be viewed as a rank-reduction technique applied to the doubles amplitudes; distance-based screening, approximation of electron pair interactions, and the introduction of orbital domains are methods for leveraging elementwise sparsity in the CC wavefunction. These methods, particularly in conjunction with the (T) correction, have found widespread application as perhaps the most efficient way to apply CC theory to the ground states of realistic chemical systems.[73] Still, it remains a challenge to achieve similar performance gains with these methods for electronic excited states and molecular response properties. While there is progress along these lines, a performant implementation of a PNO-based methods for response properties has yet to be developed.[74-77]

When sparsity in the CC wavefunction is ignored, some gains in computational efficiency are possible through low-rank approximation of the ERIs alone. The earliest work in this area by Rendell and Lee identified both the potential advantages and limitations of the introduction of DF approximations to the CCSD(T) method.[78] Many of the input/output (I/O) and storage bottlenecks can be reduced, however, it is not possible to reduce the computational complexity of the method. Over the past few years, several new massively parallel implementations of CC theory have emerged that are optimized to exploit DF representations of the ERIs.[10-14, 16, 79-80] Low-rank factorizations of the ERIs that expose additional algebraic flexibility such as PS, COS and THC offer the benefits of DF or CD for reductions in I/O and storage, but also allow the computational complexity of certain contributions to the CC amplitude equations to be reduced. Most importantly, the rate-limiting "particle-particle ladder" term in CCSD can be implemented with $O(N^5)$ scaling.[81-84] Unfortunately, low-rank factorization of the ERIs alone cannot reduce the asymptotic complexity of CCSD which remains $O(N^6)$, and the scaling of storage requirements remains $O(N^4)$. Analogous results have been demonstrated for periodic CC theory in the context of both DF and THC approximations.[85-87] In order to achieve scaling reductions in CCSD using only low-rank tensor approximations, the rank structure of the doubles amplitudes must also be considered.

In order to reduce the computational complexity of CCSD from $O(N^6)$ to $O(N^4)$, THC factorizations of both the ERIs and the doubles amplitudes are needed. There have been several realizations of such methods. We developed a THC-based CCSD method using least-squares THC (LS-THC) to approximate the doubles amplitudes.[88] This work demonstrated that chemical accuracy in the correlation energy could be achieved with an $O(N^4)$ scaling implementation of CCSD. Benedikt *et al.* developed a coupled-cluster doubles (CCD) method based on a four-way CANDECOMP/PARAFAC (CP) factorization[89-90] of the doubles amplitudes at each iteration.[91] Later, Scuseria and co-workers developed an implementation of CCSD that fully optimized a THC factorization of the doubles amplitudes at each iteration.[92] Although these works were successful proofs of concept establishing reduced scaling and accuracy, the software implementations of these methods have yet to progress beyond the pilot stage and their full practical potential was never realized. These methods had some formal issues as well. The usual CCSD method has $O(N^4)$ amplitude equations to be solved with $O(N^4)$ unknowns. In low rank CCSD methods, there are only $O(N^2)$ unknowns. This leads to questions of how to define the amplitude equations in these low rank CCSD methods.

Recently, we introduced the rank-reduced coupled-cluster (RR-CC) method.[93-94] In this approach, the doubles amplitudes are globally projected into a reduced space. By using the CC energy functional as a starting point for the theory, well-defined equations can be derived for the compressed doubles amplitudes (now a matrix rather than a fourth-order tensor). We have extended this method to excited states via equation-of-motion[95] (EOM) CC and developed an optimized GPU implementation.[93, 96] In our work to this point, we have compressed the CCSD wavefunction using truncated eigendecompositions of the doubles amplitudes. We have found that the combination of this compression of the doubles amplitudes and DF-like factorizations of the ERIs enables computations on systems 50% larger than what would be possible with DF alone – molecular systems with about 150 atoms and 1500 basis functions can be treated with rank-reduced CCSD (RR-CCSD).[96] This method is primarily limited by its $O(N^6)$ scaling and the need to construct some intermeditates that require $O(N^4)$ storage. Our RR-CC methodology has some similarities to earlier work by Kinoshita, Hino, and Bartlett who decomposed the doubles amplitudes asymmetrically using singular value decomposition (SVD).[97-98] Their later work on CC methods including triples corrections bears even more similarities to our RR-CC method both in the form of the low-rank factorization and the definition of the equations used to update the factor matrices.[98] More recently, these ideas have been revived in impressive work by Lesiuk applying Tucker-3 decompositions[99] to the triples amplitudes in full and approximate coupled-cluster singes, doubles and triples (CCSDT) methods.[100-101] What remains is to combine the ideas and formal grounding of RR-CC with THC factorizations of both the ERIs and the doubles amplitudes.

In this work, we introduce THC approximation of the ERIs and the doubles amplitudes to the RR-CCSD method. We show that there are essentially no changes required to the RR-CC formalism to make it compatible with a THC-factorized wavefunction. Further, we develop a performant graphical processing unit (GPU) implementation of the method that exhibts $O(N^4)$ scaling with system size. We demonstrate that it is scalable to molecular systems with approximately 250 atoms and 2500 basis functions using a single NVIDIA DGX A100 workstation. We test the computational efficiency of the implementation as well as its accuracy in describing both absolute and relative energies.

## II. Theory

Throughout this work, we will adhere to the following convention for labeling of different classes of indices.

- i,j,k,l: Indexes occupied molecular orbitals. Range is $N_\mathrm{o}$.

- a,b,c,d: Indexes virtual molecular orbitals. Range is $N_\mathrm{v}$.

- p,q,r,s: Indexes all molecular orbitals. Range is $N_\mathrm{MO}$.

- $\mu,\nu,\lambda,\sigma$: Indexes atomic orbitals. Range is $N_\mathrm{AO}$.

- A,B,C,D: Indexes the auxiliary index of the Cholesky decomposed ERIs. Range is $N_\mathrm{CD}$.

- I,J,K,L: Indexes the auxiliary index of THC factorized ERIs. Range is $N_\mathrm{THC}$.

- P,Q,R,S: Indexes the auxiliary index of the rank-reduced doubles amplitudes. Range is $N_\mathrm{RR}$.

- W,X,Y,Z: Indexes the auxiliary index of the THC factorized doubles amplitudes. Range is $N_\mathrm{aux}$.

For typical applications of THC-CCSD, we assume that the relative sizes of these indices are:

$$N_\mathrm{o} < N_\mathrm{v} < N_\mathrm{MO} \approx N_\mathrm{AO} < N_\mathrm{CD} \approx N_\mathrm{THC} \approx N_\mathrm{RR} \approx N_\mathrm{aux} \,. \tag{1}$$

### II.a. Rank-Reduced Coupled-Cluster Theory

In rank-reduced coupled-cluster singles and doubles (RR-CCSD),[94] we begin from the coupled-cluster energy functional,

$$E_\mathrm{CC} = \langle \Phi_0 | (1 + \hat{\Lambda}) e^{-\hat{T}} \hat{H} e^{\hat{T}} | \Phi_0 \rangle \tag{2}$$

where $\hat{T}$ is a cluster excitation operator and $\hat{\Lambda}$ is a de-excitation operator. Next, we assert the following form for the doubles amplitudes.

$$t_{ij}^{ab} \equiv \sum_{PQ} U_{ia}^P T^{PQ} U_{jb}^Q \tag{3}$$

Here, $T^{PQ}$ is a matrix containing a compressed representation of the doubles amplitudes and the $U_{ia}^{P}$ tensors are projectors that serve to transform between the full-rank and compressed spaces. After redefining the cluster excitation operator and de-excitation operators to include this low-rank factorization,

$$\hat{T} = \sum_{ia} t_i^a \hat{a}_a^\dagger \hat{a}_i + \frac{1}{4} \sum_{ijab} \left[ \sum_{PQ} U_{ia}^P T^{PQ} U_{jb}^Q \right] \hat{a}_a^\dagger \hat{a}_b^\dagger \hat{a}_j \hat{a}_i \tag{4}$$

$$\hat{\Lambda} = 1 + \sum_{ia} \lambda_a^i \hat{a}_i^\dagger \hat{a}_a + \frac{1}{4} \sum_{ijab} \left[ \sum_{PQ} U_{ia}^P \Lambda^{PQ} U_{jb}^Q \right] \hat{a}_i^\dagger \hat{a}_j^\dagger \hat{a}_b \hat{a}_a \tag{5}$$

the $t$-amplitude equations can be defined by taking partial derivatives of the coupled-cluster energy functional with respect to the amplitudes of $\hat{\Lambda}$ and setting them equal to zero. While the definitions of the energy and singles amplitudes remain the same as usual in CCSD, the equation defining the doubles amplitudes becomes:

$$0 = U_{ia}^P U_{jb}^Q \langle \Phi_{ij}^{ab} | e^{-\hat{T}} \hat{H} e^{\hat{T}} | \Phi_0 \rangle \tag{6}$$

This is a projection of the usual CCSD amplitude equations into a low-rank space defined by the $U_{ia}^{P}$ tensors. A similar approach can be taken to define the $\hat{\Lambda}$ amplitudes if molecular properties are desired; after solving for the $t$-amplitudes, the corresponding lambda equations can be defined by setting the partial derivatives of the coupled-cluster energy functional with respect to the amplitudes of $\hat{T}$ to zero. Solving the lambda equations will result in a coupled-cluster energy functional that is stationary with respect to variation in both the $\hat{T}$ and $\hat{\Lambda}$ amplitudes. This provides a starting point for the evaluation of molecular properties. The RR-CCSD approach provides rigorously defined amplitude equations with the same number of equations and unknowns. Extensions to excited states and response properties are straightforward.[93-94]

### II.b. Tensor HyperContraction of Doubles Amplitudes

We wish to achieve a THC factorization of the doubles amplitudes of the form:

$$t_{ij}^{ab} \equiv \sum_{PQ} \sum_{WX} y_i^W y_a^W \tau^{PW} T^{PQ} \tau^{QX} y_j^X y_b^X$$

$$(7)$$

This can be obtained through a CP factorization of the projectors, $U_{ia}^P$, used in RR-CCSD.

$$U_{ia}^P \approx \sum_W y_i^W y_a^W \tau^{PW}$$

$$(8)$$

The usual projectors of RR-CCSD are semi-unitary owing to their origin as a subset of the eigenvectors of a set of doubles amplitudes.

$$\delta_{PQ} = \sum_{ia} U_{ia}^P U_{ia}^Q$$

$$(9)$$

However, once the projectors have been approximated, they lose their semi-unitary property.

$$S^{PQ} = \sum_{ia} \sum_{WX} y_i^W y_a^W \tau^{PW} y_i^X y_a^X \tau^{QX}$$

$$(10)$$

Here, the only properties guaranteed of $S^{PQ}$ are that it is symmetric and positive definite. However, we can define an orthogonalizer of the CP-approximated projectors (here, via symmetric orthogonalization).

$$\tilde{U}_{ia}^P = \sum_Q [S^{-1/2}]^{PQ} \sum_\Theta y_i^W y_a^W \tau^{QW}$$

$$(11)$$

After orthogonalization, these CP-factorized projectors recover the semi-unitary properties of the usual RR-CC projectors.

$$\delta_{PQ} = \sum_{ia} \tilde{U}_{ia}^P \tilde{U}_{ia}^Q$$

$$(12)$$

Importantly, the orthogonalizing transformation can be applied independently to one of the factor matrices.

$$\tilde{\tau}^{PW} = \sum_Q [S^{-1/2}]^{PQ} \tau^{QW}$$

(13)

Because this transformation can be applied without 'pinning' any of the factor matrices together, the projectors can be written in a form that is simultaneously CP-factorized and semi-unitary.

$$\tilde{U}_{ia}^P = \sum_W y_i^W y_a^W \tilde{\tau}^{PW}$$

(14)

As a consequence of Eqns. 12 and 14, the RR-CCSD methodology described in Reference 94, can be straightforwardly adapted to accommodate THC factorizations of the doubles amplitudes through substitution of the dense $U_{ia}^P$ projectors with their CP-factorized counterparts, $\tilde{U}_{ia}^P$.

### II.c. Obtaining Low-Rank Factorizations of Projectors

A compact factorization of the $U_{ia}^P$ projectors is desired for efficiency of the THC-CCSD method. A naïve CP decomposition can be applied to these tensors, however, this leads to a larger apparent rank than necessary because all entries of $U_{ia}^P$ are treated equivalently. When this tensor is viewed as a matrix with a compound $ov$ index, each row of this matrix has unit norm; however, the importance of each of these rows in describing the CC wavefunction is not equivalent. Therefore, we need to develop a strategy for factorizing this tensor that accounts for the relative importance of the entries. Since the $U_{ia}^P$ projectors are obtained as the eigenvectors of a set of doubles amplitudes, we have the eigenvalues readily available to quantify the importance of each row of the projector. We use the eigenvalues in a weighted least-squares optimization to solve for the CP factors.

$$\min_{\mathbf{x,y},\tau} \left\| \epsilon_P \left[ U_{ia}^P - \sum_W y_i^W y_a^W \tau^{PW} \right] \right\|_F^2$$

(15)

Here, the residual is scaled by the corresponding eigenvalue, $\epsilon_p$, and each contribution to the objective function is weighted its square, therefore the eigenvalues need not be positive. This optimization can be performed using standard alternating least squares (ALS) algorithms[102] with minor modification to incorporate the weights (see Appendix I for additional details). For example, the update rule for the $y_i^W$ factor becomes:

$$y_i^W = \sum_{aP} \epsilon_P^2 U_{ia}^P \sum_X y_a^X \tau^{PX} [\sum_b y_b^X y_b^W \sum_Q \epsilon_Q^2 \tau^{QX} \tau^{QW}]^{-1}$$

(16)

The update for the $y_a^W$ factor can be inferred from Eqn. 16 to and the update for $\tau^{PW}$ remains unchanged from the unweighted case. Empirically, we find that the introduction of the weights allows a factorization of the projectors with equivalent rank for both the truncation of the eigendecomposition of the doubles amplitudes as well as the CP decomposition to provide a suitable approximation of the projector for accurate predictions of the CCSD correlation energy (see Sec. V for extensive numerical tests of this claim).

### II.d. THC Factorizations of the Electron Repulsion Integrals

In addition to the THC factorization of the doubles amplitudes, we introduce a THC factorization of the electron repulsion integrals (ERIs).[34]

$$(ia|jb) \equiv \sum_{IJ} x_i^I x_a^I z^{IJ} x_j^J x_b^J$$

(17)

There have been many techniques developed for obtaining such THC decompositions of the ERIs by us and others.[34-35, 103-106] Any of these decompositions can be applied in the THC-CCSD method, however, compact decompositions are preferred. In the case of CC-based methods, the computational expense of solving the amplitude equations to obtain the correlation energy is quite high. The cost of the entire CC computation increases rapidly when excited states and/or properties (i.e. solving the so-called lambda equations) are also sought. Therefore, a cost-benefit analysis of the expense associated with obtaining the THC factorization of the ERIs and the rank of the decomposition favors obtaining lower rank at the cost of a more expensive THC factorization. This should be contrasted with THC-MP2 where the cost of the THC factorization impacts the overall computation time more significantly. Notice that specifically the *ovov*-type ERIs are presented in Eqn. 17. This is the only class of integral for which a THC factorization is required to reduce the overall scaling of the CCSD method to $O(N^4)$. While alternative factorizations exist that involve THC factorizations of other classes of integrals (i.e. *oooo*, *vvvv*, etc.), the THC-CCSD algorithm we develop in this work requires only the *ovov*-type ERIs in THC factorized form.

Here, we construct a THC factorization in the MP2 natural orbital (NO) basis using the orbital occupations as weights in the objective function.[81] Further, we use a Cholesky decomposition of the ERIs to approximate the ERIs used in the THC factorization. The origin of the factor matrices (i.e. the **x** and **z** matrices) will be discussed shortly.

$$O_{\text{LS-THC}} = \min_{\mathbf{z}} \left\| n_i n_a n_j n_b \left[ \sum_A L_{ia}^A L_{jb}^A - \sum_{IJ} x_i^I x_a^I z^{IJ} x_j^J x_b^J \right] \right\|_F^2 \tag{18}$$

where the subscript $F$ denotes the Frobenius norm. This is a weighted LS-THC factorization of the ERIs.[35] As has been noted previously,[35] this is equivalent to a somewhat simpler objective function:

$$O_{\text{LS-CD-THC}} = \min_{\xi} \left\| n_i n_a \left[ L_{ia}^A - \sum_I x_i^I x_a^I \xi^{AI} \right] \right\|_F^2 \tag{19}$$

where $z^{IJ}$ and $\xi^{AI}$ determined from Eqns. 18 and 19, respectively, are related by:

$$z^{IJ} = \sum_A \xi^{AI} \xi^{AJ} \tag{20}$$

This is a weighted least-squares Cholesky-decomposed THC factorization of the ERIs.[35]

In the present work, we solve for the CP factorization of the Cholesky decomposition of the ERIs in the MP2 NO basis including weights from the orbital occupations. Note that the Cholesky tensors are already weighted by the square root of diagonal elements of the ERIs, so there is no need to apply additional weighting for the auxiliary index $A$. With the above considerations, we use

$$O_{\text{PF-CD-THC}} = \min_{\mathbf{x}, \xi} \left\| n_i n_a \left[ L_{pq}^A - \sum_I x_i^I x_a^I \xi^{AI} \right] \right\|_F^2$$

(21)

As our objective function to determine the factor matrices. This is similar to the PARAFAC THC (PF-THC) methods that have been previously developed.[34] Note that this minimization is performed in the MP2 NO basis to allow the weights to be introduced. The final step of the ALS optimization is the determination of $\xi^{AI}$ – as in Eqn. 19 – thus the final factorization is equivalent to LS-THC (Eqn. 18) using the $x_i^I$ and $x_a^I$ factor matrices obtained from the minimization in Eqn. 21. Once the ALS optimization is complete, we transform the resulting matrices back to the canonical MO basis from the MP2 NO basis. A final note is that we perform the THC factorization of the doubles amplitudes prior to formation of the ERI factorization. The THC factorized form of the MP2 doubles amplitudes (i.e. the initial guess of the CCSD doubles amplitudes) can be used to construct the MP2 one-particle density matrix in $O(N^4)$ time. This allows the MP2 NO basis to be accessed without the usual $O(N^5)$ scaling of the canonical MP2 method (see Appendix II).

### II.e. Definition and Scope of the THC-CCSD Method

The RR-CCSD method is defined by the low-rank approximation of the doubles amplitudes (i.e. Eqn. 3) and amplitude equations derived from the coupled-cluster energy functional (Eqn. 2

). In this method, the ERIs can be obtained from any source: exact ERIs, density fitting, Cholesky, pseudospectral, or THC approximations. Similarly, the projectors can be obtained from any source. In practice, we find that using the eigenvectors of MP2 or MP3 doubles amplitudes is fruitful for defining the $t$-amplitudes. However, we stress that it is the form of the approximate doubles amplitudes – not the source of the projectors – that defines the RR-CCSD method.

The THC-CCSD method is similarly defined by the THC factorization of the approximate doubles amplitudes.

$$t_{ij}^{ab} \equiv \sum_{PQ} \tilde{U}_{ia}^P T^{PQ} \tilde{U}_{jb}^Q = \sum_{PQ} \sum_{WX} y_i^W y_a^W \underbrace{\tilde{\tau}^{PW} T^{PQ} \tilde{\tau}^{QX}}_{T^{WX}} y_j^X y_b^X \tag{22}$$

Here, it is possible to define the working variables in analogy to RR-CCSD using the semi-unitary basis (i.e. $T^{PQ}$), or to define the working variables in the THC auxiliary basis, $T^{WX}$. We recommend the former for numerical stability and ease of implementation. In either case, the equations defining THC-CCSD are derived from the coupled-cluster energy functional. As was the case for RR-CC methods, the THC-CC methods are not defined by the source of the projectors. Similarly, they are not defined by the approach taken to reach THC factorizations of the doubles amplitudes (i.e. Eqn. 22). In the present work, we describe CP decompositions of the projectors as one way to obtain such a THC factorization (Eqn. 8), however other possibilities exist and those that produce factorizations of the form defined in Eqn. 22 can be used in the context of our THC-CC method.

Finally, in the present work we are using a THC factorization of the ERIs. While the THC-CCSD method itself requires only a THC factorization of the amplitudes, the use of THC factorized ERIs offers the most algebraic flexibility and the most potential for scaling reductions. We have previously shown that by combining the THC factorized amplitudes with THC factorized ERIs, the asymptotic scaling of CCSD can be reduced from $O(N^6)$ to $O(N^4)$. Here, we use a CP decomposition of the Cholesky decomposed ERIs obtained by a weighted ALS optimization in the MP2 natural orbital basis. Again, we must stress that the neither the asymptotic scaling of the implementation or the representation of the ERIs defines the THC-CCSD method.

There are many remaining research questions that need to be answered in order to fully optimize the RR-CC and THC-CC methods. The optimal source of the $U_{ia}^P$ projectors needs to be determined along with efficient algorithms for forming those projectors. For THC-CC, optimized algorithms for obtaining CP decompositions of the projector need to be developed. Alternatively, methods that directly obtain the THC factorization of the doubles amplitudes (Eqn. 22) could be considered. Finally, the optimal method for obtaining THC decompositions of the ERIs in the context of THC-CC is also unknown. As these questions are answered, the RR-CC and THC-CC methods will be further refined.

## III. Implementation Details

We have developed an implementation of the THC-CCSD amplitude equations optimized for a single compute node with multiple GPUs. In our algorithm, we do not require that any 4-index quantities (i.e. $t_{ij}^{ab}$) be explicitly constructed or stored. A limited number of 3-index arrays are formed including the projectors, $U_{ia}^P$, and Cholesky vectors, $L_{\mu\nu}^A$. These arrays are stored in the CPU memory and are transferred piecewise to the GPU as needed. The 2-index arrays describing the ERIs (i.e. the THC factorization of $(ia \,|\, jb)$, Equation 17) and the doubles amplitudes,

$$t_{ij}^{ab} \equiv \sum_{WX} y_i^W y_a^W T^{WX} y_j^X y_b^X$$

(23)

are assumed to be replicated in the GPU memory. For some of the largest systems considered, the largest of these 2-index arrays is the core tensor of the THC ERI factorization, $z^{IJ}$, requires less than 1 GB of memory to store (for a system with 2480 basis functions and 400 occupied orbitals). These assumptions are reasonable for up to roughly 3000 basis functions at which point the storage of the Cholesky vectors (with a cutoff of $10^{-4} E_h$) approaches 500 GB. Scalability to systems of this size represents approximately a 2x increase over RR-CCSD and a 3x increase over canonical CCSD computations. Since these methods scale as $O(N^6)$, the corresponding RR-CCSD and CCSD calculations would be roughly 64x and 700x more demanding, respectively, than what is currently possible with those methods.

In the following presentation of our THC-CCSD algorithm, we follow the notation of Reference 107 as closely as possible. The use of a $T_1$ transformation of the Hamiltonian is utilized throughout the work. The $T_1$ transformation of the ERIs and Cholesky vectors is denoted by a hat.

$$(pq|\hat{r}s) \approx \sum_A \hat{L}_{pq}^A \hat{L}_{rs}^A$$

(24)

Here, the MO coefficient matrices are modified to include contributions from the $t_1$-amplitudes.

$$\hat{C}_{\mu a}^{\mathrm{p}} = C_{\mu a} - \sum_i t_i^a C_{\mu i}$$

(25)

$$\hat{C}_{\mu i}^{\mathrm{h}} = C_{\mu i} + \sum_a t_i^a C_{\mu a}$$

(26)

The Cholesky vectors are then transformed using this modified basis.

$$\hat{L}_{pq}^A = \sum_{\mu\nu} \hat{C}_{\mu p}^{\mathrm{p}} \hat{C}_{\nu q}^{\mathrm{h}} L_{\mu\nu}^A$$

(27)

This allows the $T_1$-transformed ERIs to be directly contructed from the transformed Cholesky vectors. It is important to note that the usual 8-fold permutational symmetry of the ERIs is lost when the $T_1$-transformation is applied – only a 2-fold bra-ket symmetry remains. It is necessary, however, to deviate from the factorization and grouping of terms presented in Reference 107 to fully exploit the algebraic flexibility introduced by the THC representation of doubles amplitudes and ERIs.

In Algorithm 1, we present a factorized implementation of the so-called ladder diagrams.

$$\sigma^{XX'} = \sum_{ijab} \tilde{U}_{ia}^X \tilde{U}_{jb}^{X'} \left[ \sum_{cd} t_{ij}^{cd} (ac\hat{|}bd) + \sum_{kl} t_{kl}^{ab} (ki\hat{|}lj) - \sum_{cl} t_{il}^{cb} (ac\hat{|}lj) - \sum_{dk} t_{kj}^{ad} (ki\hat{|}bd) \right] \tag{28}$$

These contributions are contained in $\Omega^A$, $\Omega^B$, and $\Omega^C$ of Reference [107]. By grouping these terms together the mixed particle-hole ladder terms can be evaluated as a byproduct of the particle-particle and hole-hole ladder terms. This reduces the number of floating point operations required and the transfer of the Cholesky vectors to the GPU. This operation is parallelized over GPUs by splitting the $A$ index of the Cholesky vectors to parallelize both the floating point operations and memory transfer. Formally, this parallelization is nearly perfect with only the initial formation of the $o^{XY}$ and $v^{XY}$ being performed in serial and the reduction of the $\sigma^{XX'}$ array requiring memory transfer from each GPU. The rate limiting step of this algorithm scales as $O(N_{CD}N_{aux}^3)$, that is linearly in the length of the Cholesky vectors and cubically in the rank of the THC factorization of the doubles amplitudes. These rate limiting steps are performed as matrix multiplications using NVIDIA's cuBLAS libraries.

In Algorithm 2, we include contributions from some of the ring diagrams in $\Omega^D$ and the bare $T_1$-transformed ERIs from $\Omega^F$.

$$\sigma^{XX'} = \sum_{ijab} \tilde{U}_{ia}^X \tilde{U}_{jb}^{X'} \left[ (ai\hat{|}bj) + \sum_{ck} [2t_{ik}^{ac} - t_{ik}^{ca}](bj\hat{|}kc) + \sum_{ck} [2t_{jk}^{bc} - t_{jk}^{cb}](ai\hat{|}kc) \right] \tag{29}$$

These terms rely on common intermediates formed by the contraction of the projectors with the $T_1$-transformed Cholesky vectors.

$$\bar{A}^{XA} = \sum_{ia} \tilde{U}_{ia}^X \hat{L}_{ai}^A$$

(30)

Since we again parallelize over the $A$ index of the Cholesky vector, the **A** matrix of Equation 30 is formed a vector at a time. The contributions to the amplitude equations from Algorithm 2 are somewhat less expensive than those from Algorithm 1 scaling at worst $O(N_o N_{CD} N_{aux}^2)$. Note that the operations involving double summations (lines 3, 4, 5, 6, 8, 9, 10, 12 and 13) are performed as two separate operations; the order of operations is chosen to minimize the number of floating point operations assuming that $N_o < N_v < N_{AO} < N_{aux}$.

In Algorithm 3, we present a factorization of the remaining contributions to $\Omega^C$ that are linear in $T_2$.

$$\sigma^{XX'} = -\sum_{ijab} \tilde{U}_{ia}^X \tilde{U}_{jb}^{X'} \sum_{ck} \left[ t_{ik}^{ac}(kj|bc) + t_{jk}^{bc}(ki|ac) \right]$$

(31)

These contributions scale as $O(N_v N_{CD} N_{aux}^2)$. We present Algorithms 1, 2, and 3 separately for the sake of clarity. However, in our software implementation, the operations inside these three loops over the Cholesky vectors are combined and performed simultaneously. This minimizes the memory transfer between host and device and also allows for the reuse of some common intermediates.

The most challenging contributions to evaluate arise from the contributions of $\hat{T}_2^2$ to $\Omega^A$ and $\Omega^C$. These two terms are the first two places where THC factorizations of the ERIs are used. Without THC factorizations of both the ERIs and the doubles amplitudes, it is not possible to reduce the scaling of these terms below O(N$^6$). In Algorithms 4 and 5, we present an O(N$^5$) scaling factorization that relies on matrix multiplications for all the rate determining steps.

$$\sigma^{XX'} = \sum_{ijab} \tilde{U}_{ia}^X \tilde{U}_{jb}^{X'} \sum_{kcld} t_{ij}^{cd} t_{kl}^{ab} (kc|ld) \tag{32}$$

$$\sigma^{XX'} = \sum_{ijab} \tilde{U}_{ia}^X \tilde{U}_{jb}^{X'} \sum_{kcld} t_{ik}^{cb} t_{lj}^{ad} (lc|kd) \tag{33}$$

Algorithms 4 and 5 evaluate the contributions to the amplitude equations from Equations 32 and 33, respectively. These algorithms follow the canonical strategy for evaluating these terms. However, they leverage THC factorizations to reduce the overall scaling and memory footprint. In Algorithm 4, Lines 3 – 5 construct $t_{ij}^{cd}$ explicitly. Lines 6 – 8 contract those doubles amplitudes with the ERIs: $X_{ij}^{kl} = \sum_{cd} t_{ij}^{cd} (kc|ld)$. Lines 9 – 11 contract that intermediate array with the second set of doubles amplitudes: $\sigma_{ij}^{ab} = \sum_{kl} X_{ij}^{kl} t_{kl}^{ab}$. Finally, Lines 12 – 13 project the residual into the low-rank space using the $\tilde{U}_{ia}^P$ projectors. These arrays are formed in slices, so no arrays with more than 2-indices are held in memory. The algebraic flexibility of the THC factorization allows these exchange-like contractions to be implemented with reduced scaling. Algorithm 5 follows a similar pattern to what was described for Algorithm 4. The rate limiting steps of Algorithm 4 and 5 scale as $O(N_o^2 N_v M^2)$ where $M = \max(N_{aux}, N_{THC})$. The advantage of this algorithm is that it can leverage the floating point performance available on the GPU and, in practice, we find it to be quite efficient for smaller molecules.

An O(N$^4$) implementation of Equations 32 and 33 is presented in Algorithm 6. Here, both contributions are evaluated simultaneously (Equation 34) using custom CUDA kernels to perform the rate limiting operations.

$$\sigma^{XX'} = \sum_{ijab} \tilde{U}_{ia}^X \tilde{U}_{jb}^{X'} \sum_{kcld} \left[ t_{ij}^{cd} t_{kl}^{ab} (kc|ld) + t_{ik}^{cb} t_{lj}^{ad} (lc|kd) \right]$$

(34)

This algorithm starts by constructing 3-index arrays; Line 2 reduces the THC form of the ERIs to something that resembles the psuedospectral approximation. Looping over the THC ERI index allows a 3-index array to be constructed by accumulating a series of Khatri-Rao products and Hadamard products. Line 11 is rate limiting and implemented with a custom CUDA kernel. The overall memory usage of this algorithm is controlled by blocking over the $X$ index. This limits the memory usage associated with the large $C_i^{XY}$, $D_a^{XY}$, and $H^{XYZ}$ arrays. It should be noted, however, that not every operation depends on the $X$ index (see Line 10, for example), so as the number of blocks increases, so does the amount of redundant work performed. Similarly, the complete $B_{ia}^I$ tensor must be transferred to the GPU for each block. Nevertheless, we find this algorithm to become optimal for systems with more than roughly 1000 basis functions. This is the rate-limiting step in a O($N^4$)-scaling implementation of THC-CCSD with Line 11 scaling as $O(N_{THC} N_{aux}^3)$.

In Algorithm 7, we evaluate the contributions from $\hat{T}_2^2$ to $\Omega^D$ and explicitly remove one spurious contribution that is usually cancelled through the evaluation of $\Omega^C$ according to Ref. [107].

$$\sigma^{XX'} = \frac{1}{4} \sum_{ijab} \tilde{U}_{ia}^X \tilde{U}_{jb}^{X'} \sum_{klcd} \left[ (2t_{ik}^{ac} - t_{ik}^{ca}) \left[ 2(kc|ld) - (kd|lc) \right] \left( t_{jl}^{bd} - t_{jl}^{db} \right) + t_{ik}^{ca} (kd|lc) t_{jl}^{db} \right]$$

(35)

This algorithm can be viewed in two steps. First, the projectors are contracted with the doubles amplitudes (Lines 1 – 8).

$$R_{ia}^X = \sum_{jb} \tilde{U}_{jb}^X \left( 2t_{ij}^{ab} - t_{ij}^{ba} \right)$$

(36)

Then, this intermediate is contracted with the THC ERIs to form the complete contribution to the residual (Lines 9 – 22).

$$\sigma^{XX'} = \frac{1}{4} \sum_{ijab} R_{ia}^X R_{jb}^{X'} \left[ 2(ia|jb) - (ib|ja) \right]$$

(37)

Notice that there is an additional term evaluated to removed the spurious contribution; this term reuses many of the intermediates constructed in the solution of Equations 36 and 37. The formation of the $R_{ia}^X$ intermediate is blocked over the $X$ index both as a way to parallelize the formation (by assigning blocks to different GPUs) and as a way to control memory usage on the GPU. In the subsequent contraction with the THC ERIs, the algorithm loops over the $X$ index to reduce memory usage and to parallelize the memory transfer back to the GPU. This algorithm contains steps that scale as $O(N_o N_{\mathrm{aux}}^3)$ and $O(N_o N_{\mathrm{aux}} N_{\mathrm{THC}}^2)$; these rate limiting steps are performed as matrix multiplications.

In Algorithm 8, the contributions to the singles residual from $\boldsymbol{\Omega}^G$ and $\boldsymbol{\Omega}^H$ are evaluated along with two intermediates that will be used to simplify the evaluation of $\boldsymbol{\Omega}^E$.

$$\sigma_{ia} = \sum_{kcd} t_{ik}^{cd} \left[ 2(kd\hat{|}ac) - (kc\hat{|}ad) \right] - \sum_{klc} t_{kl}^{ac} \left[ 2(lc\hat{|}ki) - (li\hat{|}kc) \right]$$

(38)

$$\xi_{kj} = \sum_{dem} t_{jm}^{de} \left[ 2(me|kd) - (md|ke) \right]$$

(39)

$$\xi_{bc} = - \sum_{lmd} t_{lm}^{db} \left[ 2(ld|mc) - (lc|md) \right]$$

(40)

The evaluation of these terms is parallelized by blocking over the $A$ index of the Cholesky vector (as was done in Algorithms 1 – 3). The rate limiting operations scale as $O(N_o N_{\mathrm{CD}} N_{\mathrm{aux}}^2)$ and can be performed as matrix multiplications. A parallel reduction is required to assemble the contributions to $\xi_{kj}$ and $\xi_{bc}$ from each GPU.

Using the intermediates defined in Equations 39 and 40, the remaining contributions to the doubles residual from $\Omega^E$ are evaluated following Algorithm 9.

$$\sigma^{XX'} = \sum_{ijab} \tilde{U}_{ia}^X \tilde{U}_{jb}^{X'} \left[ \sum_c \left( t_{ij}^{ac} + t_{ji}^{ca} \right) \left( \hat{F}_{bc} + \xi_{bc} \right) - \sum_k \left( t_{ik}^{ab} + t_{ki}^{ba} \right) \left( \hat{F}_{kj} + \xi_{kj} \right) \right] \tag{41}$$

The work to evaluate these contributions scales only as $O(N^3)$. Finally, for completeness, we present the $\Omega^I$ and $\Omega^J$ contributions to the singles residual in Algorithm 10.

$$\sigma_{ia} = \hat{F}_{ai} + \sum_{kc} \left( 2t_{ik}^{ac} - t_{ik}^{ca} \right) \hat{F}_{kc} \tag{42}$$

These terms are trivial to evaluate and scale as $O(N^3)$ in our implementation.

## IV. Computational Details

The GPU-accelerated implementation of the THC-CCSD method described in this work is implemented in the TERACHEM electronic structure package.[108-111] The ERIs are obtained from a fully-pivoted incomplete Cholesky decomposition. In all computations, the frozen core approximation is applied to the $1s$ orbitals of second row elements. Timings presented in this work were performed on a single NVIDIA DGX-1 (eight NVIDIA V100 GPUs and two 20-core 2.2GHz Intel Xeon E5-2698 CPUs) or a single NVIDIA DGX-A100 (eight NVIDIA A100 GPUs and two 64-core 2.25 GHz AMD EPYC 7742 CPUs).

## V. Results and Discussion

To test the scalability of the THC-CCSD algorithm and compare to existing CCSD and RR-CCSD implementations, we perform a series of CCSD/TZVP computations on water clusters ranging in size from 15 atoms (155 basis functions) to 240 atoms (2480 basis functions). The results of this comparision are shown in Figure 1. As expected the CCSD and RR-CCSD iterations scale as $O(N^6)$. Due to the need to store many arrays of $N_o^2 N_v^2$ size, the CCSD algorithm is limited to roughly 100 atoms and 1000 basis functions. The RR-CCSD algorithm requies only a single $N_o^2 N_v^2$ array to be stored and therefore can be applied to slightly larger systems (up to 150 atoms and 1500 basis functions). Owing to both the reduced asymptotic scaling and the lack of explicit evaluation or storage of 4-index arrays, the THC-CCSD method can be applied to systems with up to 240 atoms and 2500 basis functions. Empirically, we find the scaling of the THC-CCSD algorithms to be somewhat lower than the expected $O(N^4)$ and $O(N^5)$ complexity. There are two primary factors contributing to the lower observed scaling. First, the total computation involves many operations with less than $O(N^4)$ or $O(N^5)$ complexity. Since these operations take a non-negligible amount of time, the overall scaling appears lower than what is expected in the asymptotic limit. Second, the computational efficiency of these algorithms increases with system size; this accounts for the increase in computational time for smaller system sizes that does not follow a simple power law.

We find that the THC-CCSD algorithms perform $t$-amplitude iterations faster than both CCSD and RR-CCSD for all but the smallest system considered. Crossover between THC-CCSD and (RR-)CCSD occurs between 155 and 310 basis functions. The $O(N^4)$-scaling THC-CCSD algorithm, however, does not become optimal until systems with between 775 and 930 basis functions are reached. This is because the rate limiting steps in the $O(N^5)$ THC-CCSD algorithm are matrix multiplications and are performed using NVIDIA's cuBLAS libraries. The high computational efficiency of these operations leads to a lower multiplicative prefactor and a more efficient algorithm for smaller system sizes. The computational time associated with the $O(N^4)$ algorithm is dominated by a custom CUDA kernel with a low ratio of floating point operations to memory operations. Ideally, one would implement a THC-CCSD algorithm that used heuristics to choose between the more efficient of the $O(N^4)$ and $O(N^5)$ factorizations.

In our development of the THC-CCSD algorithm, we focused on strategies that enable efficient multi-GPU parallelization. In Figure 2, we characterize the parallel efficiency of the THC-CCSD iterations with up to 8 GPUs. The test systems in this case were clusters of water molecules computed using a TZVP basis set and a $10^{-4}$ cutoff on MP2-based projectors. We find that the parallel efficiency over two GPUs is excellent and exceeds 90% for systems with as few as 10 water molecules. With four GPUs, the 90% threshold for parallel efficiency is reached for 20 water molecules; with eight GPUs, 90% efficiency is not reached until 45 water molecules are included. The parallel efficiency exceeds what we obtained in our earlier work on the GPU acceleration of CCSD.[12] The key to obtaining high parallel efficiency in THC-CCSD is the reduced amount of data transfer relative to the number of floating point operations performed. The loops over the number of Cholesky vectors that appear in Algorithms 1, 2, 3 and 8 expose parallelism in the THC-CCSD amplitude equations that can be easily exploited. By assigning each GPU a range of indices, the work (both the floating point operations and data transfer) is distributed almost perfectly over the GPUs. It is also noteworthy that 90% parallel efficiency is reached with 8 GPUs for systems that are roughly half of the maximum size accessible with this algorithm. We expect that the parallel efficiency will increase for larger molecular systems.

While efficient algorithms to perform the $t$-amplitude iterations are important, the THC-CCSD method also includes a significant amount of overhead to construct the factorized representations of the doubles amplitudes and ERIs necessary to obtain that efficiency. In Figure 3, we consider the total time required to solve for the CCSD correlation energy using the canonical CCSD method as well as THC-CCSD. With CCSD, the only significant overhead present is the Cholesky decomposition of the ERIs. As a result, the time spent in the $t$-amplitude iterations dominates the cost of the computation for systems with more than 300 basis functions.

We also note that our implementation of CCSD uses the $T_1$-transformation of the Hamiltonian described in Ref. 107. Therefore, we consider the AO to MO transformation of the ERIs to be part of the $t$-amplitude iterations, since it incorporates contributions from the $T_1$ amplitudes and changes at each iteration.

The THC-CCSD method introduces three significant additional sources of overhead: the formation of the projectors, the THC factorization of the projector and the THC factorization of the ERIs. Importantly, each of these algorithms scales as $O(N^4)$ and all are less computationally demanding than the rate limiting contribution to the THC-CCSD amplitude equations. Therefore, we expect that in the large-scale limit the cost of solving the amplitude equations will dominate the total cost of the THC-CCSD method. Here, we show that crossover between the total time to perform CCSD and THC-CCSD computations occurs at around 300 basis functions. However, for systems of this size, the total time spent in the $t$-amplitude iterations is less than what is required to perform the Cholesky decomposition and THC factorization of the ERIs. For systems with more than 750 basis functions, the $t$-amplitude iterations become the dominant contribution to the cost of the THC-CCSD method; for systems with more than 1000 basis functions, the $t$-amplitude iterations take more than 50% of the total computation time. (see Figure 4).

The relative amount of time spent in the $t$-amplitude iterations during THC-CCSD computations indicates that there is room to further optimize the implementation. In particular, replacing the Cholesky decomposed ERIs with density fitted ERIs or improving the efficiency of the Cholesky decomposition algorithm could offer a significant speedup for smaller molecules. Obtaining the THC factorization of the projectors and ERIs using a GPU-accelerated ALS algorithm does not seem to be problematic. This approach introduces only a modest amount of overhead for larger molecules. The formation of the MP2-based projector (i.e. finding the eigenvectors of the MP2 amplitudes) is also a significant source of overhead in THC-CCSD computations. The algorithm we use to define the projectors is outlined in Ref. 96. While the rate limiting steps in the formation of the MP2 projectors are GPU accelerated, the large matrix diagonalizations are, at present, performed on the CPU. Moving a larger fraction of this algorithm to the GPU could improve performance. Finally, we note that for the largest systems considered, the time to evaluate the THC-CCSD energy is dominated almost entirely by the $t$-amplitude iterations.

We have previously shown that accurate CCSD correlation energies can be obtained by using projectors defined as the eigenvectors of MP3 doubles amplitudes. The formation of these projectors is costly and scales as $O(N^5)$. To obtain accurate excitation energies from EOM-CCSD, we have shown that it is sufficient to use projectors for the $t$-amplitudes defined from MP2 amplitudes. As such, the timings shown in Figures 1 – 4 should be understood to be indicative of what would be required to construct a similarity transformed Hamiltonian as a starting point for calculating EOM-CCSD excitation energies. In the present case of CCSD, however, we will use MP3-based projectors (and their THC factorizations) to test the accuracy of THC-CCSD. To make this method more practical, better or alternative algorithms to solve for accurate projectors need to be developed. A final note is that RR-CCSD or THC-CCSD using MP2-based projectors could be understood as a new coupled-cluster method instead of an approximation to CCSD. In future work, we will characterize the performance of RR-CCSD based on MP2 projectors to assess its advantages compared to methods of similar cost.

To put the performance of the THC-CCSD method in context with the state-of-the-art in conventional coupled-cluster computations, we compare the THC-CCSD implementation reported in the present work to our GPU-accelerated CCSD program,[12] the CCSD implementation of Krylov and coworkers based on Cholesky decomposed ERIs (CD-CCSD),[16, 80] and several massively parallel implementations of density-fitted CCSD (DF-CCSD) that have been developed over the past few years.[10-11] The results of this comparision are presented in Table 1. A first observation is that on average THC-CCSD iterations are 10× to 30× faster than the corresponding CD-CCSD iterations using our GPU accelerated implementation depending on the thresholds used in the projector formation. Some of this difference is a result of the use of more capable hardware (the CD-CCSD computations used the older NVIDIA DGX-1 while the THC-CCSD computations used the NVIDIA DGX-A100), however, most of the speedup is a result of the low-rank approximations to the ERIs and amplitudes as well as the factorization of the amplitude equations. Compared to CD-CCSD on a single CPU node, the THC-CCSD iterations are 1000× to 3000× faster. Perhaps the more interesting comparision is to massively parallel implementations of DF-CCSD. In those cases, it appears that 700 to 800 CPU nodes are required to perform one DF-CCSD iteration in the same time as one THC-CCSD iteration on a single NVIDIA DGX-A100 using a $10^{-6}$ threshold for the projectors. With a looser threshold of $10^{-4}$, the number of CPUs nodes needed to match the performance of THC-CCSD increases to somewhere between 2500 and 2850. This indicates that the combination of rank-reduction, the factorized amplitude equations and an optimized GPU implementation allows one to obtain results with a single NVIDIA DGX-A100 that would otherwise require the dedicated use of a very large computer cluster. Of course, much of the increased efficiency is a result of the approximations introduced by THC-CCSD. To better understand the tradeoff between computational efficiency and accuracy in THC-CCSD a thorough error analysis is required.

The next question is whether or not the THC-CCSD method is capable of producing accurate correlation energies with compact factorizations of the doubles amplitudes and ERIs.

Here, we will hold the rank of the CP decomposition used to obtain the THC factorization equal to the rank of the underlying eigendecomposition or Cholesky decomposition of the projectors and ERIs, respectively. This leads to modest memory requirements to store the factorized projectors and ERIs as well as limiting the cost to perform the ALS optimization of the CP factor matrices. We test the accuracy of THC-CCSD using MP3-based projectors for water clusters (Figure 5), alanine helices (Figure 6) and a series of carbon clusters (Figure 7).

Here, we define chemical accuracy in the correlation energy to be 0.1% error; that is, 1 mE$_h$ error per 1 E$_h$ of correlation energy. Errors below 0.01% error are assumed to be suffiently small for the vast majority of practical applications. We find that compact THC factorizations using MP3-based projectors provide correlation energies at or below chemical accuracy with cutoffs on the MP3 eigenvalues at $10^{-3}$. Reducing this cutoff to $10^{-4}$ typically provides better than 0.01% error. Counterintuitively, the performance of THC-CCSD is worst for the water clusters, however, this appears to be a result of error cancellation between the rank reduction of the doubles amplitudes (i.e. application of the MP3 projectors) and the THC factorization of those projectors. The excellent performance for the larger covalently bonded systems (Figures 6 and 7) are encouraging. We note that the relative errors do not appear to grow with system size. This is an essential characteristic of a robust low-rank approximation.

Finally, we apply the THC-CCSD method to predict the relative energy of 27 tetraalanine conformers (Figure 8). This tests the suitability of the THC-CCSD approximations to practical applications in quantum chemistry. An analysis of the errors is presented in Table 2. We find that THC-CCSD using MP2 projectors is capable of predicting the conformational energies with accuracy of 0.10 ± 0.13 kcal mol$^{-1}$. Using MP3 projectors, this error is reduced to -0.02 ± 0.04 kcal mol$^{-1}$. This reduction in error with the increase of fidelity in the projector indicates that this is the primary source of error in the THC-CCSD method, not the THC factorization of the ERIs. Interestingly, despite the marked increase in accuracy moving from MP2 to MP3 projectors, the errors are not significantly reduced by decreasing the cutoff thresholds on the projectors. This indicates that it is the THC factorization of the projectors that is limiting the accuracy of the method. Moving forward, it may be necessary to develop better optimization algorithms for these THC factors. Encouragingly, the THC-CCSD method provides sufficient accuracy to resolve the energy differences between these tetraalanine conformers using MP2 projectors, suggesting that the more costly MP3 projectors may not be merited in most chemical applications.

## VI. Conclusions

The introduction of THC approximations of the doubles amplitudes and ERIs to RR-CCSD enables the asymptotic scaling of the method to be reduced from $O(N^6)$ to $O(N^4)$. The underlying RR-CCSD formalism remains nearly unchanged, since the THC factorization of the doubles amplitudes is introduced through CP factorization of the projectors. This allows the factorized projector to be orthogonalized while retaining all the algebraic flexibility of the THC factorization (Eq. 14). This also means that the compressed doubles amplitudes – the working variables in RR-CCSD and THC-CCSD – remain exactly the same size. THC factorization of the ERIs is only required to reduce the complexity of certain terms in the doubles residual resulting from $\left\langle \Phi_{ij}^{ab} \middle| \left[\left[\hat{V}, \hat{T}_2\right], \hat{T}_2\right] \middle| \Phi_0 \right\rangle$. Our GPU implementation of the THC-CCSD method exhibits the expected $O(N^4)$ scaling and is applicable to molecular systems with up to 250 atoms and 2500 basis functions. For systems of this size, it is possible to perform the complete THC-CCSD computation in less than one day on a single NVIDIA DGX A100 workstation. This demonstrates how effective the THC-CCSD approximations can be at reducing the computational cost of the CCSD method.

The accuracy in the correlation energy achieved by the underlying RR-CCSD method is preserved in THC-CCSD. The accuracy is determined primarily by the quality of the projector with MP2 projectors being sufficiently accurate for many chemical applications and MP3-based projectors being available when higher accuracy is required. Following our findings with RR-EOM-CCSD, we expect that MP2 projectors for the ground state will be more than sufficient in a forthcoming THC-EOM-CCSD method. At present, overall $O(N^4)$ scaling is only possible when MP2-based projectors are used. Due to the simple mathematical form of the MP2 doubles amplitudes, it is straightforward to obtain their eigenvectors with effort scaling as $O(N^4)$. The MP3 amplitudes, however, require a significant amount of additional effort. In recent work by Lesiuk, an algorithm is presented to solve directly for the MP3-based projectors with effort scaling as $O(N^5)$.[112] Although this would dominate the cost of the THC-CCSD computation, it would still enable the overall scaling of CCSD to be reduced. Projector formation in RR-CC methods is an area of active research and is the key to unlocking increased accuracy in these methods. However, we stress that MP2-projectors provide sufficient accuracy for many (perhaps most) practical applications of an $O(N^4)$-scaling THC-CCSD method.

The obvious extension of the THC-CCSD method is to excited states via EOM-CCSD. We have previously developed the formalism required for RR-EOM-CCSD; the introduction of THC factorizations into that method is straightforward and follows along lines similar to the present work. Our expectations are that an implementation of THC-EOM-CCSD can be developed that exhibits $O(N^4)$ asymptotic scaling with excitation energies accurate to roughly 0.01 eV compared to the canonical result. We expect the cost of determining each excitation energy to be roughly twice the cost of determining the ground state correlation energy, but the limits on the system size that can be treated will be similar to that of the ground state (because the storage requirements are effictively identical). In our estimation, this should make it routine to perform EOM-CCSD computations on molecular systems with up to 150 atoms and possible to perform EOM-CCSD computations on systems with up to 250 atoms. Work to develop such an implementation is underway in our laboratory.

**Acknowledgements**

This material is based upon work supported by the U.S. Department of Energy, Office of Science, Basic Energy Sciences, Chemical Sciences, Geosciences and Biosciences Division.

**Appendix.**

*I. GPU acceleration of alternating least squares CANDECOMP/PARAFAC factorizations*

In this work, we must perform efficient CP factorizations of 3-index arrays. Here, we seek a decomposition of the form:

$$A_{ijk} \approx \sum_n X_i^n Y_j^n Z_k^n$$

$$(43)$$

In this section, we use $i$, $j$, and $k$ to index the 3-index array and $n$ for the auxiliary index of the CP factorization. To that end, we have developed a GPU-accelerated implementation of the ALS algorithm that parallelizes over multiple GPUs on a single node. In ALS, two of the factor matices are held constant and the other is determined via least-squares optimization; this optimization procedure is repeated for each of the factor matrices. The optimization of the factor matrices is performed iteratively until convergence is reached (where one ALS iteration for a 3-index array consists of three least-squares optimizations). This algorithm is amenable to parallelization over multiple GPUs by exploiting the fact that the linear least squares procedure can be trivially parallelized over each of the right hand sides. This can be seen from the ALS update rule for the $X$ factor matrix.

$$\sum_{n'} S_{nn'} X_i^{n'} = R_i^n$$

(44)

where,

$$S_{nn'} = \left( \sum_j Y_j^n Y_j^{n'} \right) \left( \sum_k Z_k^n Z_k^{n'} \right)$$

(45)

and

$$R_i^n = \sum_{jk} A_{ijk} Y_j^n Z_k^n$$

(46)

An update of the $i$-th row of the $X$ factor matrix is independent of all other rows of $X$. In our approach, we replicate the factor matrices ($X$, $Y$, and $Z$) on each GPU in addition to forming the $S$ matrix (Eqn. 45) redundantly. The $A$ array can be blocked over the $i$ index so that it fits in the GPU memory and blocks of the array can be transmitted to the GPU. The contraction in Eqn. 46 (that is, to form the right-hand side of Eqn. 44) is performed directly using custom GPU kernels to avoid the formation of additional 3-index arrays (to conserve GPU memory). Subsequently, the linear system in Eqn. 44 is solved on the GPU by the preconditioned conjugate gradient method using the diagonal of $S$ as the preconditioner. The blocks of the $X$ factor matrix are then transferred back to the host CPU. We note that it is often the case for CCSD-type computations that the GPUs have sufficient memory to allow the $A$ array to be replicated three times in the distributed memory of the GPUs. In that case, the blocks of the $A$ arrays only need to be transferred to the GPUs once and the overall memory transfers between host and device are limited to updates of the factor matrices after each solution of Eqn. 44. This algorithm can be easily extended to multiple nodes each with multiple GPUs, however, the parallel efficiency will be limited by the redundant formation and storage of the $S$ matrix (Eqn. 45).

## *II. Reduced-scaling formation of the MP2 one-particle density matrix*

The use of the MP2 natural orbital basis provides a practical approach to improve the accuracy and compactness of THC factorizations of the ERIs. However, the formation of the one-particle density matrices scales canonically as $O(N^5)$.

$$D_{ij} = -2 \sum_{kab} t_{ik}^{ab} \left[ 2t_{jk}^{ab} - t_{jk}^{ba} \right]$$

(47)

$$D_{ab} = 2 \sum_{ijc} t_{ij}^{ac} \left[ 2t_{ij}^{bc} - t_{ji}^{bc} \right]$$

(48)

To reduce this scaling, THC factorizations of the MP2 doubles amplitudes can be introduced. Note that formation of the MP2-based projectors, THC factorization of these projectors and the subsequent formation of the THC factorized MP2 doubles amplitudes (which will serve as the initial guess for the THC-CCSD amplitude equations) each scale as $O(N^4)$. Inserting these THC factorizations in Eqns. 47 and 48 results in expressions for the MP2 density matrices that can be evaluated in $O(N^4)$.

$$D_{ij} = -2 \sum_{kab} \sum_{WX} y_i^W y_a^W T^{WX} y_k^X y_b^X \sum_{YZ} \left[ 2y_j^Y y_a^Y T^{YZ} y_k^Z y_b^Z - y_j^Y y_b^Y T^{YZ} y_k^Z y_a^Z \right]$$

(49)

$$D_{ab} = 2 \sum_{ijc} \sum_{WX} y_i^W y_a^W T^{WX} y_j^X y_c^X \sum_{YZ} \left[ 2y_i^Y y_b^Y T^{YZ} y_j^Z y_c^Z - y_j^Y y_b^Y T^{YZ} y_i^Z y_c^Z \right]$$

(50)

The Coulomb-like contributions to the density matrices can be formed efficiently by first defining two new matrices.

$$o^{WX} = \sum_i y_i^W y_i^X$$

(51)

$$v^{WX} = \sum_i y_a^W y_a^X$$

(52)

With these matrices, it becomes clear how to implement the Coulomb-like contributions with $O(N^3)$ scaling.

$$D_{ij} \leftarrow -4 \sum_W y_i^W \left( \sum_Y y_j^Y \left( v^{WY} \left( \sum_X T^{WX} \left( \sum_Z \left( o^{XZ} v^{XZ} \right) T^{YZ} \right) \right) \right) \right) \tag{53}$$

$$D_{ab} \leftarrow 4 \sum_W y_a^W \left( \sum_Y y_b^Y \left( o^{WY} \left( \sum_X T^{WX} \left( \sum_Z \left( o^{XZ} v^{XZ} \right) T^{YZ} \right) \right) \right) \right) \tag{54}$$

These expressions should be evaluated starting from the innermost parenthesis and working outwards. Where summations appear, matrix multiplications are performed; otherwise, Hadamard products are performed. The exchange-like contractions make use of similar intermediates, however the factorization of these contributions are somewhat more involved.

$$D_{ij} \leftarrow 2 \left( \sum_Y y_j^Y \left( \sum_X v^{XY} \left( \sum_Z T^{YZ} \left( o^{XZ} \left( \sum_a y_a^Z \left( \sum_W T^{WX} \left( y_i^W y_a^W \right) \right) \right) \right) \right) \right) \right) \tag{55}$$

$$D_{ab} \leftarrow -2 \left( \sum_Y y_b^Y \left( \sum_X o^{XY} \left( \sum_Z T^{YZ} \left( v^{XZ} \left( \sum_i y_i^Z \left( \sum_W T^{WX} \left( y_i^W y_a^W \right) \right) \right) \right) \right) \right) \right) \tag{56}$$

Again, working from the innermost parenthesis outwards leads to an O(N$^4$) scaling evaluation of the MP2 one-particle density matrix.

**Table 1.** Comparison of the wall time required for one CCSD iteration for several parallel CCSD codes and our GPU-accelerated implementation of THC-CCSD.

| Software | Method | Nodes | Total Cores | GPUs | Iteration Time (s) |
|---|---|---|---|---|---|
| AATT, 6-311+G**, 60 atoms, 966 AOs | | | | | |
| QChem[a] | CD-CCSD/FNO | 1 | 12 | 0 | 133200 |
| QChem[b] | CD-CCSD | 1 | 8 | 0 | 110880 |
| TeraChem[c] | CD-CCSD | 1 | 8 | 8 V100 | 960 |
| TeraChem[d] | THC-CCSD ($10^{-4}$) | 1 | 8 | 8 A100 | 34 |
| TeraChem[d] | THC-CCSD ($10^{-5}$) | 1 | 8 | 8 A100 | 55 |
| TeraChem[d] | THC-CCSD ($10^{-6}$) | 1 | 8 | 8 A100 | 86 |
| Beta carotene, mTZ, 96 atoms, 1032 AOs | | | | | |
| FHI-aims[e] | DF-CCSD | 32 | 640 | 0 | 3780 |
| FHI-aims[e] | DF-CCSD | 512 | 10240 | 0 | 360 |
| TeraChem[c] | CD-CCSD | 1 | 8 | 8 V100 | 1680 |
| TeraChem[d] | THC-CCSD ($10^{-4}$) | 1 | 8 | 8 A100 | 72 |
| TeraChem[d] | THC-CCSD ($10^{-5}$) | 1 | 8 | 8 A100 | 138 |
| TeraChem[d] | THC-CCSD ($10^{-6}$) | 1 | 8 | 8 A100 | 276 |
| GC-dDMP-B, 6-311++G**, 63 atoms, 1042 AOs | | | | | |
| NWChem[f] | CCSD | 1100 | 1100 | 0 | 4320 |
| NWChem[f] | CCSD | 20000 | 20000 | 0 | 780 |
| MPQC[g] | DF-CCSD | 64 | 1024 | 0 | 2580 |
| TeraChem[c] | CD-CCSD | 1 | 8 | 8 V100 | 1500 |
| TeraChem[d] | THC-CCSD ($10^{-4}$) | 1 | 8 | 8 A100 | 58 |
| TeraChem[d] | THC-CCSD ($10^{-5}$) | 1 | 8 | 8 A100 | 102 |
| TeraChem[d] | THC-CCSD ($10^{-6}$) | 1 | 8 | 8 A100 | 210 |
| Pentacene Dimer, aug-cc-pVDZ, 72 atoms, 1264 AOs | | | | | |
| MPQC[h] | DF-CCSD | 64 | 4096 | 0 | 11520 |
| TeraChem[c] | CD-CCSD | 1 | 8 | 8 V100 | 2400 |
| TeraChem[d] | THC-CCSD ($10^{-4}$) | 1 | 8 | 8 A100 | 90 |
| TeraChem[d] | THC-CCSD ($10^{-5}$) | 1 | 8 | 8 A100 | 168 |
| TeraChem[d] | THC-CCSD ($10^{-6}$) | 1 | 8 | 8 A100 | 318 |

a. Reference 80. One node with two 6-core Intel Xeon X5675 3.0 GHz CPUs. The frozen natural orbital (FNO) approximation reduced the number of virtuals from 830 to 551.

b. Reference 16. One node with two 4-core Intel Xeon E5-2623 3.0 GHz CPUs.

c. Reference 12. NVIDIA DGX-1 with 8 NVIDIA V100 GPUs.

d. This work. NVIDIA DGX-A100 with 8 NVIDIA A100 GPUs. The $O(N^5)$ scaling THC-CCSD implementation is used for these computations.

e. Reference 11. Each node has two 8-core Intel Xeon E5-2670 2.6 GHz CPUs.

f. Reference 15. Each node has two 8-core AMD 6276 2.3 GHz Interlagos CPUs.

g. Reference 10. Each node has two 8-core Intel Xeon E5-2670 2.6 GHz CPUs.

h. Reference 10. Each node has one 64-core Intel Xeon Phi 7230 1.3 GHz CPU.

**Table 2.** Relative errors of THC-CCSD/TZVP for tetraalanine conformational energies using low-rank factorization of the cluster amplitudes compared to canonical CCSD.

| Method (Projector) | Average Error | Standard Deviation | RMS Error | Maximum Absolute Error |
|---|---|---|---|---|
| Tetraalanine Conformational Energies (kcal mol$^{-1}$)[a] | | | | |
| THC-CCSD (MP2 $\varepsilon=10^{-4}$) | 0.086 | 0.085 | 0.120 | 0.303 |
| THC-CCSD (MP2 $\varepsilon=10^{-5}$) | 0.093 | 0.090 | 0.128 | 0.305 |
| THC-CCSD (MP2 $\varepsilon=10^{-6}$) | 0.109 | 0.082 | 0.135 | 0.313 |
| THC-CCSD (MP3 $\varepsilon=10^{-3}$) | -0.022 | 0.030 | 0.037 | 0.107 |
| THC-CCSD (MP3 $\varepsilon=10^{-4}$) | -0.023 | 0.028 | 0.036 | 0.074 |

[a] 27 conformers of tetraalanine from Ref. 113.

1:    Form: $o^{XY} = \sum_i y_i^X y_i^Y$

2:    Form: $v^{XY} = \sum_a y_a^X y_a^Y$

3:    **for** $A$ **do**

4:        Transfer: $L_{\mu\nu}^A$ on the CPU to $l_{\mu\nu}$ on the GPU

5:        Form: $l_{ij} = \sum_{\mu\nu} \hat{C}_{\mu i}^{\mathrm{p}} \hat{C}_{\nu j}^{\mathrm{h}} l_{\mu\nu}$

6:        Form: $\bar{A}_i^X = \sum_k y_k^X l_{ki}$

7:        Form: $\bar{B}^{XY} = \sum_i \bar{A}_i^X y_i^Y$

8:        Form: $\bar{C}^{XY} = v^{XY} \bar{B}^{XY}$

9:        Form: $l_{ab} = \sum_{\mu\nu} \hat{C}_{\mu a}^{\mathrm{p}} \hat{C}_{\nu b}^{\mathrm{h}} l_{\mu\nu}$

10:        Form: $\tilde{A}_c^X = \sum_a y_a^X l_{ac}$

11:        Form: $\tilde{B}^{XY} = \sum_c \tilde{A}_c^X y_c^Y$

12:        Form: $\tilde{C}^{XY} = o^{XY} \tilde{B}^{XY}$

13:        Form: $C^{XY} = \bar{C}^{XY} - \tilde{C}^{XY}$

14:        Form: $D^{XY'} = \sum_Y C^{XY} T^{YY'}$

15:        Accumulate: $\sigma^{XX'} \leftarrow \sum_{Y'} D^{XY'} C^{X'Y'}$

16:    **end for**

17:    Transfer: $\sigma^{XX'}$ from GPU to CPU

**Algorithm 1.** Pseudocode for the evaluation of contributions to Omega A, B and C from $\hat{T}_2$. This algorithm assumes the **y** and **T** matrices are resident in the GPU memory along with the molecular orbital coefficients. This algorithm can be parallelized over multiple GPUs by patitioning the loop over the $A$ index (line 3) into equal segments and assigning each to one of the GPUs.

1:    **for $A$ do**

2:        Transfer: $L^A_{\mu\nu}$ on the CPU to $l_{\mu\nu}$ on the GPU

3:        Form: $l_{ai} = \sum_{\mu\nu} \hat{C}^{\mathrm{p}}_{\mu a} \hat{C}^{\mathrm{h}}_{\nu i} l_{\mu\nu}$

4:        Form: $l_{ia} = \sum_{\mu\nu} \hat{C}^{\mathrm{p}}_{\mu i} \hat{C}^{\mathrm{h}}_{\nu a} l_{\mu\nu}$

5:        Form: $\overline{A}^X = \sum_{ai} y^X_a y^X_i l_{ai}$

6:        Form: $\tilde{A}^X = \sum_{ia} y^X_i y^X_a l_{ia}$

7:        Form: $\tilde{B}^Y = \sum_{Y'} T^{YY'} \tilde{A}^{Y'}$

8:        Form: $\tilde{C}_{ia} = \sum_Y y^Y_i y^Y_a \tilde{B}^Y$

9:        Form: $\tilde{D}^X = \sum_{ia} y^X_i y^X_a \tilde{C}_{ia}$

10:       Form: $E^{YY'} = \sum_{kc} y^Y_c y^{Y'}_k l_{kc}$

11:       Form: $F^{YY'} = T^{YY'} E^{YY'}$

12:       Form: $G_{ia} = \sum_{YY'} y^Y_i y^{Y'}_a F^{YY'}$

13:       Form: $H^X = \sum_{ia} y^X_i y^X_a G_{ia}$

14:       Accumulate: $\sigma^{XX'} \leftarrow \overline{A}^X \overline{A}^{X'} + 2\tilde{D}^X \overline{A}^{X'} + 2\overline{A}^X \tilde{D}^{X'} - H^X \overline{A}^{X'} - \overline{A}^X H^{X'}$

15:    **end for**

16:    Transfer: $\sigma^{XX'}$ from GPU to CPU

**Algorithm 2.** Pseudocode for the evaluation of Omega D and F contributions from $\hat{T}_2$.

1:     **for $A$ do**

2:         Transfer: $L^A_{\mu\nu}$ on the CPU to $l_{\mu\nu}$ on the GPU

3:         Form: $l_{ij} = \sum_{\mu\nu} \hat{C}^{\mathrm{p}}_{\mu i} \hat{C}^{\mathrm{h}}_{\nu j} l_{\mu\nu}$

4:         Form: $l_{ab} = \sum_{\mu\nu} \hat{C}^{\mathrm{p}}_{\mu a} \hat{C}^{\mathrm{h}}_{\nu b} l_{\mu\nu}$

5:         Form: $A^{X'Y'} = \sum_{bc} y^{X'}_b y^{Y'}_c l_{bc}$

6:         Form: $B^{X'Y'} = \sum_{kj} y^{X'}_j y^{Y'}_k l_{kj}$

7:         Accumulate: $C^{X'Y'} \leftarrow A^{X'Y'} B^{X'Y'}$

8:     **end for**

9:     Form: $o^{XY} = \sum_i y^X_i y^Y_i$

10:    Form: $v^{XY} = \sum_a y^X_a y^Y_a$

11:    Form: $D^{XY'} = \sum_Y [o^{XY} v^{XY}] T^{YY'}$

12:    Form: $\sigma^{XX'} = -\sum_{Y'} D^{XY'} C^{X'Y'} - \sum_{Y'} C^{XY'} D^{X'Y'}$

13:    Transfer: $\sigma^{XX'}$ from GPU to CPU

**Algorithm 3.** Pseudocode for the evaluation of contributions to Omega C from $\hat{T}_2$. This algorithm assumes the **y** and **T** matrices are resident in the GPU memory along with the molecular orbital coefficients. This algorithm can be parallelized over multiple GPUs by patitioning the loop over the $A$ index (line 3) into equal segments and assigning each to one of the GPUs.

1:    **for** $i$ **do**

2:        **for** $j$ **do**

3:           Form: $A^{ZZ'} = y_i^Z y_j^{Z'}$

4:           Form: $B^{ZZ'} = T^{ZZ'} A^{ZZ'}$

5:           Form: $C_{cd} = \sum_{ZZ'} B^{ZZ'} y_c^Z y_d^{Z'}$

6:           Form: $D^{IJ} = \sum_{cd} C_{cd} x_c^I x_d^J$

7:           Form: $E^{IJ} = Z^{IJ} D^{IJ}$

8:           Form: $F_{kl} = \sum_{IJ} E^{IJ} x_k^I x_l^J$

9:           Form: $G^{YY'} = \sum_{kl} F_{kl} y_k^Y y_l^{Y'}$

10:          Form: $H^{YY'} = T^{YY'} G^{YY'}$

11:          Form: $I_{ab} = \sum_{YY'} H^{YY'} y_a^Y y_b^{Y'}$

12:          Form: $J^{XX'} = \sum_{ab} I_{ab} y_a^X y_b^{X'}$

13:          Accumulate: $\sigma^{XX'} \leftarrow J^{XX'} A^{XX'}$

14:        **end for**

15:    **end for**

16:    Transfer: $\sigma^{XX'}$ from GPU to CPU

**Algorithm 4.** Pseudocode for the O(N$^5$) evaluation of Omega A contributions from $\hat{T}_2^2$.

1:    **for** $j$ **do**

2:        **for** $a$ **do**

3:            Form: $A^{ZZ'} = y_a^Z y_j^{Z'}$

4:            Form: $B^{ZZ'} = T^{ZZ'} A^{ZZ'}$

5:            Form: $C_{ld} = \sum_{ZZ'} B^{ZZ'} y_l^Z y_d^{Z'}$

6:            Form: $D^{IJ} = \sum_{ld} C_{ld} x_l^I x_d^J$

7:            Form: $E^{IJ} = Z^{IJ} D^{IJ}$

8:            Form: $F_{kc} = \sum_{IJ} E^{IJ} x_c^I x_k^J$

9:            Form: $G^{YY'} = \sum_{kc} F_{kc} y_c^Y y_k^{Y'}$

10:           Form: $H^{YY'} = T^{YY'} G^{YY'}$

11:           Form: $I_{ib} = \sum_{YY'} H^{YY'} y_i^Y y_b^{Y'}$

12:           Form: $J^{XX'} = \sum_{ib} I_{ib} y_i^X y_b^{X'}$

13:           Accumulate: $\sigma^{XX'} \leftarrow J^{XX'} A^{XX'}$

14:        **end for**

15:    **end for**

16:    Transfer: $\sigma^{XX'}$ from GPU to CPU

**Algorithm 5.** Pseudocode for the O(N$^5$) evaluation of Omega C contributions from $\hat{T}_2^2$.

1:    Form: $A_{ia}^X = \sum_Y y_i^Y y_a^Y T^{XY}$ on the CPU

2:    Form: $B_{ia}^I = \sum_J x_i^J x_a^J Z^{IJ}$ on the CPU

3:    Transfer: $A_{ia}^X$ on the CPU to $A_{ia}^X$ on the GPU

4:    Form: $C_i^{XY} = \sum_a y_a^X A_{ia}^Y$

5:    Form: $D_a^{XY} = \sum_i y_i^X A_{ia}^Y$

6:    **for** $I$ **do**

7:        Transfer: $B_{ia}^I$ on the CPU to $b_{ia}$ on the GPU

8:        Form: $E^{XY} = \sum_i C_i^{XY} x_i^I$

9:        Form: $F^{XZ} = \sum_a D_a^{XZ} x_a^I$

10:       Form: $G^{YZ} = \sum_{ia} y_i^Y y_a^Z b_{ia}$

11:       Accumulate: $H^{XYZ} \leftarrow \left( E^{XY} F^{XZ} + E^{XZ} F^{XY} \right) G^{YZ}$

12:   **end for**

13:   **for** $X$ **do**

14:       Form: $I_{ia} = \sum_{YZ} y_i^Z y_a^Y H^{XYZ}$

15:       Form: $\sigma^{XY} = \sum_{ia} y_i^Y y_a^Y I_{ia}$

16:   **end for**

17:   Transfer: $\sigma^{XX'}$ from GPU to CPU

**Algorithm 6.** Pseudocode for the O($N^4$) evaluation of Omega A and C contributions from $\hat{T}_2^{\,2}$.

1:     Form: $o^{XY} = \sum_i y_i^X y_i^Y$

2:     Form: $v^{XY} = \sum_a y_a^X y_a^Y$

3:     Form: $A^{XY} = \sum_Z o^{XZ} v^{XZ} T^{YZ}$

4:     Form: $R_{ia}^X = 2\sum_Y y_i^Y y_a^Y A^{XY}$

5:     **for** $X$ **do**

6:          Form: $B^{YZ} = v^{XY} o^{XZ} T^{YZ}$

7:          Accumulate: $R_{ia}^X \leftarrow S_{ia}^X = -\sum_{YZ} y_i^Y y_a^Z B^{YZ}$

8:     **end for**

9:     Form: $C^{XI} = \sum_{ia} x_i^I x_a^I R_{ia}^X$

10:    Form: $D^{XI} = \sum_J C^{XJ} Z^{IJ}$

12:    **for** $X$ **do**

13:         Form: $E^{IJ} = \sum_{ia} x_i^I x_a^J R_{ia}^X$

14:         Form: $F^{IJ} = E^{IJ} Z^{IJ}$

15:         Form: $G_{ia}^X = \sum_{IJ} x_i^J x_a^I F^{IJ}$

17:         Form: $\tilde{E}^{IJ} = \sum_{ia} x_i^I x_a^J S_{ia}^X$

18:         Form: $\tilde{F}^{IJ} = \tilde{E}^{IJ} Z^{IJ}$

19:         Form: $\tilde{G}_{ia}^X = \sum_{IJ} x_i^J x_a^I \tilde{F}^{IJ}$

21:    **end for**

22:    Form: $\sigma^{XY} = \frac{1}{4}\left(2\sum_J C^{XJ} D^{YJ} - \sum_{ia} G_{ia}^X R_{ia}^Y + \sum_{ia} \tilde{G}_{ia}^X S_{ia}^Y\right)$

23:    Transfer: $\sigma^{XX'}$ from GPU to CPU

**Algorithm 7.** Pseudocode for the $O(N^4)$ evaluation of Omega D contributions from $\hat{T}_2^2$.

1:     **for** $A$ **do**

2:         Transfer: $L_{\mu\nu}^{A}$ on the CPU to $l_{\mu\nu}$ on the GPU

3:         Form: $l_{ij} = \sum_{\mu\nu} \hat{C}_{\mu i}^{\mathrm{p}} \hat{C}_{\nu j}^{\mathrm{h}} l_{\mu\nu}$

4:         Form: $l_{ab} = \sum_{\mu\nu} \hat{C}_{\mu a}^{\mathrm{p}} \hat{C}_{\nu b}^{\mathrm{h}} l_{\mu\nu}$

5:         Form: $l_{ia} = \sum_{\mu\nu} \hat{C}_{\mu i}^{\mathrm{p}} \hat{C}_{\nu a}^{\mathrm{h}} l_{\mu\nu}$

6:         Form: $A^{X} = \sum_{ia} y_{i}^{X} y_{a}^{X} l_{ia}$

7:         Form: $B^{XY} = \sum_{ia} y_{i}^{X} y_{a}^{Y} l_{ia}$

8:         Form: $C^{XY} = 2\delta_{XY} \sum_{Z} T^{XZ} A^{Z} - B^{XY} T^{XY}$

9:         Form: $D_{ia} = \sum_{XY} y_{i}^{Y} y_{a}^{X} C^{XY}$

10:         Accumulate: $\xi_{ij} \leftarrow \sum_{a} l_{ia} D_{ja}$

11:         Accumulate: $\xi_{ab} \leftarrow -\sum_{i} D_{ia} l_{ib}$

12:         Accumulate: $\sigma_{ia} \leftarrow \sum_{b} D_{ib} l_{ab}$

13:         Accumulate: $\sigma_{ia} \leftarrow -\sum_{j} l_{ij} D_{ja}$

14:     **end for**

15:     Transfer: $\sigma_{ia}$, $\xi_{ij}$, $\xi_{ab}$ from GPU to CPU

**Algorithm 8.** Pseudocode for the O(N$^4$) evaluation of Omega G and H.

| | | |
|---|---|---|
| 1: | Form: | $o^{XY} = \sum_i y_i^X y_i^Y$ |
| 2: | Form: | $v^{XY} = \sum_a y_a^X y_a^Y$ |
| 3: | Form: | $A^{XY} = o^{XY} v^{XY}$ |
| 4: | Form: | $B^{XY} = \sum_Z A^{XZ} T^{ZY}$ |
| 5: | Form: | $C_i^X = \left( F_{ji} + \xi_{ji} \right) y_j^X$ |
| 6: | Form: | $D_a^X = \left( F_{ab} + \xi_{ab} \right) y_b^X$ |
| 7: | Form: | $\tilde{o}^{XY} = \sum_i y_i^X C_i^Y$ |
| 8: | Form: | $\tilde{v}^{XY} = \sum_a y_a^X D_a^Y$ |
| 9: | Form: | $E^{XY} = o^{XY} \tilde{v}^{XY} - \tilde{o}^{XY} v^{XY}$ |
| 10: | Form: | $\sigma^{XY} = \sum_Z E^{XZ} B^{YZ} + \sum_Z B^{XZ} E^{YZ}$ |
| 11: | Transfer: | $\sigma^{XY}$ from GPU to CPU |

**Algorithm 9.** Pseudocode for the evaluation of Omega E using intermediate quantities from Algorithm 8.

1:  Form: $A_i^X = \sum_a \hat{F}_{ia} y_a^X$

2:  Form: $B^{XY} = \sum_i y_i^X A_i^Y$

3:  Form: $C^X = \sum_Y T^{XY} B^{YY}$

4:  Form: $D_i^X = y_i^X C^X$

5:  Form: $\sigma_{ia} \leftarrow 2 \sum_X D_i^X y_a^X$

6:  Form: $E^{XY} = T^{XY} B^{XY}$

7:  Form: $F_i^X = \sum_Y y_i^y E^{XY}$

8:  Form: $\sigma_{ia} \leftarrow - \sum_X F_i^X y_a^X$

9:  Transfer: $\sigma_{ia}$ from GPU to CPU

10: Add: $\sigma_{ia} \leftarrow \hat{F}_{ai}$

**Algorithm 10.** Pseudocode for the evaluation of Omega I and J.

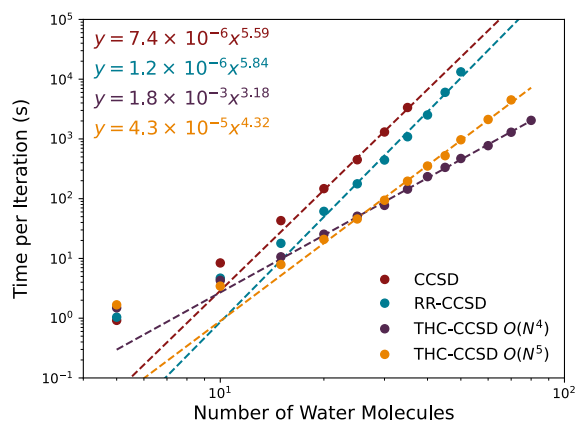

**Figure 1.** Timings of one CCSD/TZVP iteration for a series of water clusters containing between 5 and 80 water molecules using canonical CCSD, RR-CCSD and THC-CCSD algorithms. In all cases, a cutoff threshold of $10^{-4}$ $E_h$ was used during the Cholesky decomposition of the ERI tensor. For RR-CCSD and THC-CCSD, MP2 projectors are used with cutoffs on the eigenvalues of $10^{-4}$. Timings were performed using 8 NVIDIA V100 GPUs and two 20-core Intel Xeon E5-2698 CPUs clocked at 2.2 GHz.

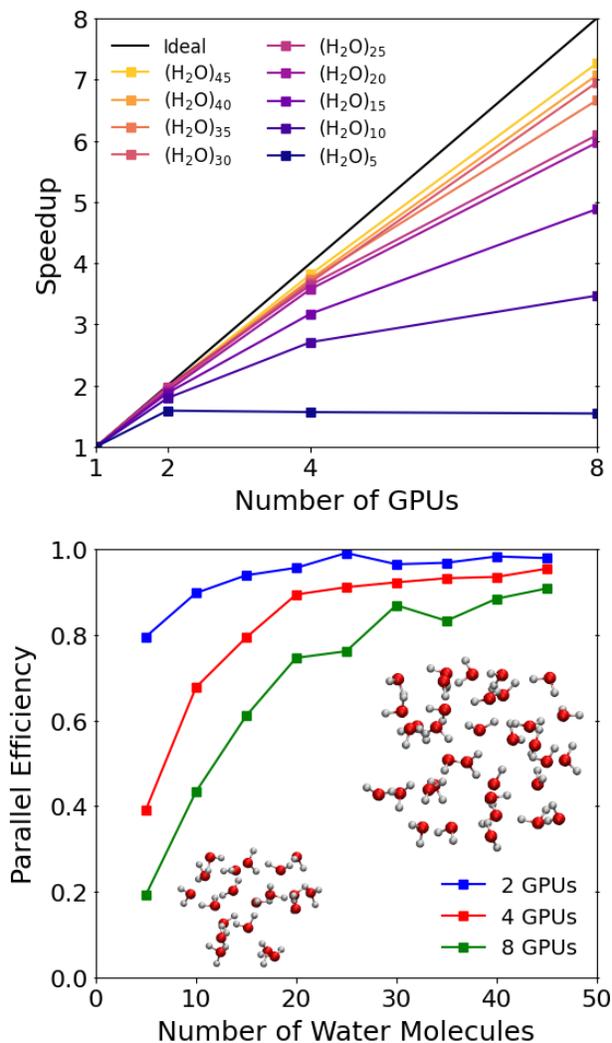

**Figure 2.** Parallel efficiency of one CCSD/TZVP iteration for a series of water clusters (ranging in size from 5 to 45 molecules). The upper panel shows the speedup obtained using multiple GPUs. The lower panel reports the parallel efficiency. The $O(N^5)$ scaling THC-CCSD algorithm was used for these computations. A cutoff threshold of $10^{-4}$ $E_h$ was used during the Cholesky decomposition of the ERI tensor. MP2 projectors were used with cutoffs on the eigenvalues of $10^{-4}$. Timings were performed using 8 NVIDIA A100 GPUs and two 64-core AMD EPYC 7742 CPUs clocked at 2.25 GHz.

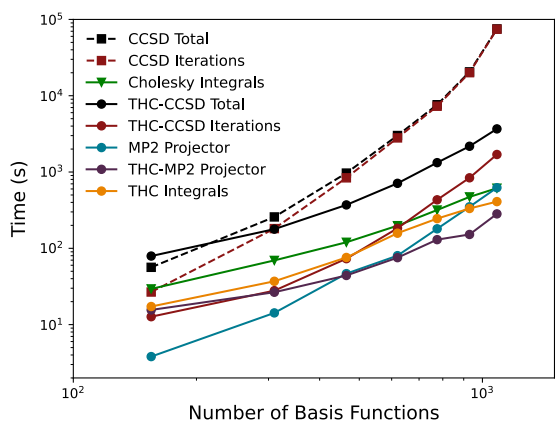

**Figure 3.** Timings of CCSD/TZVP computations for a series of water clusters (ranging in size from 5 to 35 molecules) using canonical CCSD and THC-CCSD algorithms. The $O(N^5)$ scaling THC-CCSD algorithm was used for these computations. The residual of $t$-amplitude equations was converged to $10^{-8}$. In both cases, a cutoff threshold of $10^{-4}$ $E_h$ was used during the Cholesky decomposition of the ERI tensor. For THC-CCSD, MP2 projectors were used with cutoffs on the eigenvalues of $10^{-4}$. Timings were performed using 8 NVIDIA A100 GPUs and two 64-core AMD EPYC 7742 CPUs clocked at 2.25 GHz.

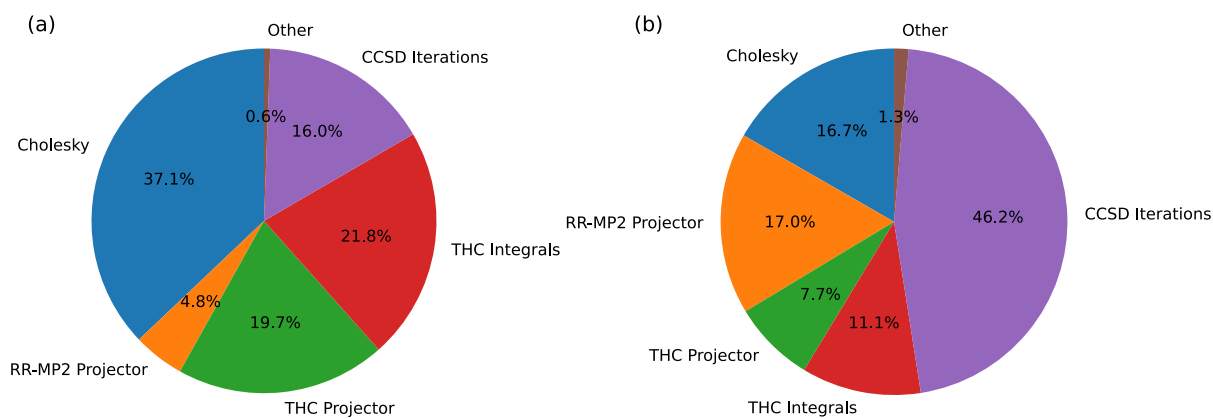

**Figure 4.** Comparison of the relative time spent in each component of a THC-CCSD/TZVP computation. Timings were performed for (a) a cluster of 5 water molecules and (b) a cluster of 35 water molecules. The $O(N^5)$ scaling THC-CCSD algorithm was used for these computations. The residual of $t$-amplitude equations was converged to $10^{-8}$. In both cases, a cutoff threshold of $10^{-4}$ $E_h$ was used during the Cholesky decomposition of the ERI tensor. For THC-CCSD, MP2 projectors were used with cutoffs on the eigenvalues of $10^{-4}$. Timings were performed using 8 NVIDIA A100 GPUs and two 64-core AMD EPYC 7742 CPUs clocked at 2.25 GHz.

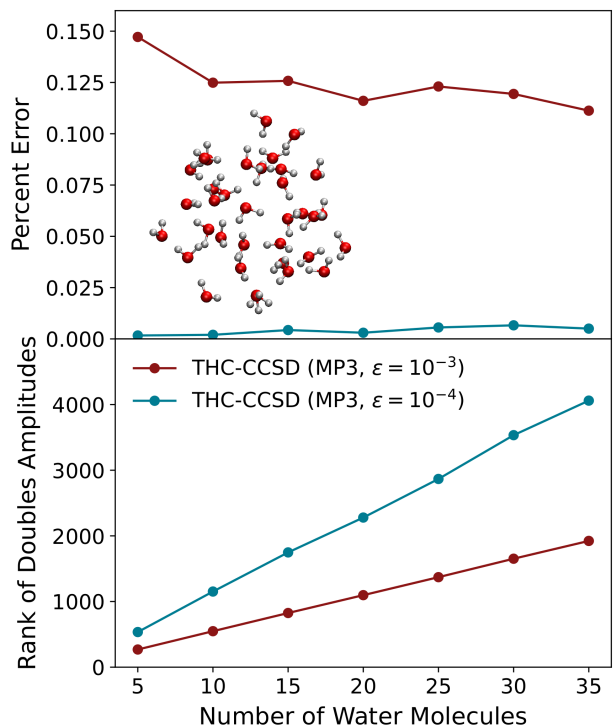

**Figure 5.** (Upper panel) Percent error in THC-CCSD/TZVP correlation energies relative to CCSD/TZVP for a series of water clusters ranging in size from 5 molecules to 35 molecules. MP3 projectors are used with cutoffs on the eigenvalues of $10^{-3}$ and $10^{-4}$. (Lower panel) Rank of the MP3 projectors. A cutoff threshold of $10^{-4}$ $E_h$ was used during the Cholesky decomposition of the ERI tensor for both CCSD and THC-CCSD. Molecular geometries were taken from Ref. 96.

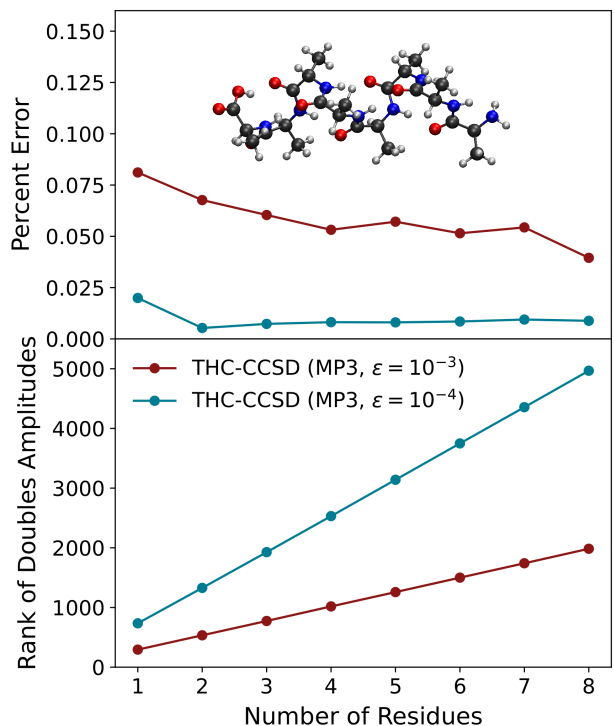

**Figure 6.** (Upper panel) Percent error in THC-CCSD/TZVP correlation energies relative to CCSD/TZVP for a series of alanine helices containing between 1 and 8 alanine residues. MP3 projectors are used with cutoffs on the eigenvalues of $10^{-3}$ and $10^{-4}$. (Lower panel) Rank of the MP3 projectors. A cutoff threshold of $10^{-4}$ $E_h$ was used during the Cholesky decomposition of the ERI tensor for both CCSD and THC-CCSD. Molecular geometries were taken from Ref. 96.

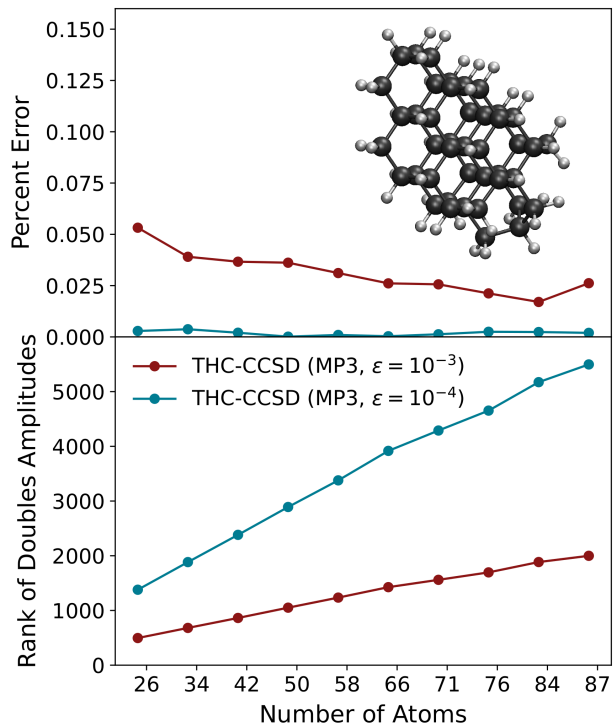

**Figure 7.** (Upper panel) Percent error in THC-CCSD/TZVP correlation energies relative to CCSD/TZVP for a series of carbon clusters. MP3 projectors are used with cutoffs on the eigenvalues of $10^{-3}$ and $10^{-4}$. (Lower panel) Rank of the MP3 projectors. A cutoff threshold of $10^{-4}$ $E_h$ was used during the Cholesky decomposition of the ERI tensor for both CCSD and THC-CCSD. Molecular geometries were taken from Ref. 96.

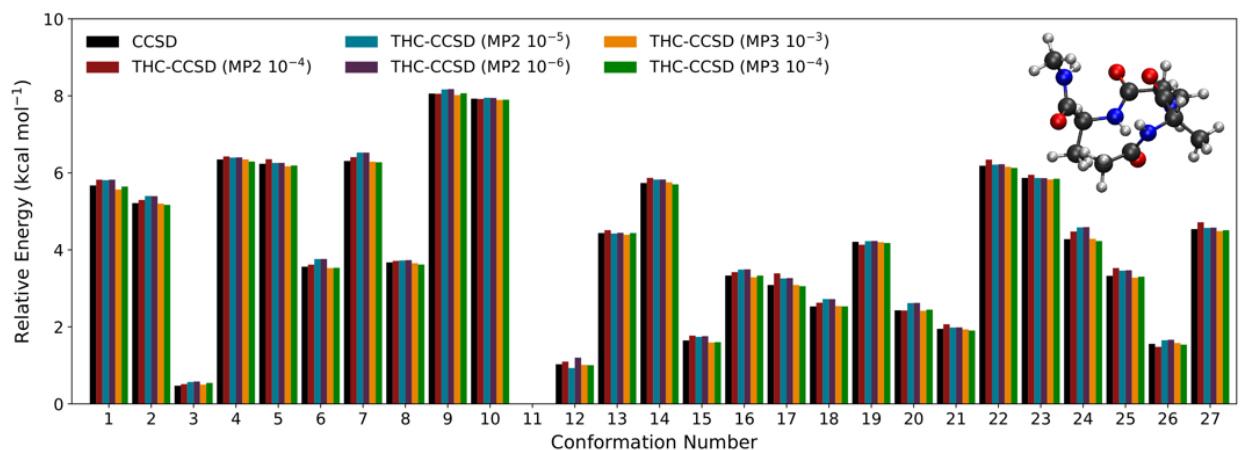

**Figure 8.** Relative energies of 27 tetraalanine conformers[113] evaluated at the CCSD/TZVP and THC-CCSD/TZVP levels of theory. THC-CCSD computations use MP2 projectors with cutoffs on the eigenvalues of $10^{-4}$, $10^{-5}$, and $10^{-6}$; MP3 projectors use cutoffs on the eigenvalues of $10^{-3}$ and $10^{-4}$. A cutoff threshold of $10^{-4}$ $E_h$ was used during the Cholesky decomposition of the ERI tensor for both CCSD and THC-CCSD.

# References

1.   Shavitt, I.; Bartlett, R. J., *Many-body methods in chemistry and physics : MBPT and coupled-cluster theory*. Cambridge University Press: Cambridge ; New York, 2009; p xiii, 532 p.

2.   Cizek, J., On Correlation Problem in Atomic and Molecular Systems . Calculation of Wavefunction Components in Ursell-Type Expansion Using Quantum-Field Theoretical Methods. *Journal of Chemical Physics* **1966**, *45* (11), 4256-+.

3.   Čížek, J., On the Use of the Cluster Expansion and the Technique of Diagrams in Calculations of Correlation Effects in Atoms and Molecules. *Advances in Chemical Physics* **1969**, 35-89.

4.   Čížek, J.; Paldus, J., Correlation problems in atomic and molecular systems III. Rederivation of the coupled-pair many-electron theory using the traditional quantum chemical methodst. *International Journal of Quantum Chemistry* **1971**, *5* (4), 359-379.

5.   Koch, H.; Jorgensen, P., Coupled Cluster Response Functions. *Journal of Chemical Physics* **1990**, *93* (5), 3333-3344.

6.   Bartlett, R. J.; Purvis, G. D., Many-Body Perturbation-Theory, Coupled-Pair Many-Electron Theory, and Importance of Quadruple Excitations for Correlation Problem. *International Journal of Quantum Chemistry* **1978**, *14* (5), 561-581.

7.   Pople, J. A.; Krishnan, R.; Schlegel, H. B.; Binkley, J. S., Electron Correlation Theories and Their Application to Study of Simple Reaction Potential Surfaces. *International Journal of Quantum Chemistry* **1978,** *14* (5), 545-560.

8.   Bartlett, R. J., Many-Body Perturbation Theory and Coupled Cluster Theory for Electron Correlation in Molecules. *Annual Review of Physical Chemistry* **1981,** *32* (1), 359-401.

9.   Raghavachari, K.; Trucks, G. W.; Pople, J. A.; Head-Gordon, M., A 5th-Order Perturbation Comparison of Electron Correlation Theories. *Chem Phys Lett* **1989,** *157* (6), 479-483.

10.   Peng, C.; Calvin, J. A.; Valeev, E. F., Coupled-cluster singles, doubles and perturbative triples with density fitting approximation for massively parallel heterogeneous platforms. *International Journal of Quantum Chemistry* **2019**, *119* (12), e25894.

11.   Shen, T.; Zhu, Z.; Zhang, I. Y.; Scheffler, M., Massive-Parallel Implementation of the Resolution-of-Identity Coupled-Cluster Approaches in the Numeric Atom-Centered Orbital Framework for Molecular Systems. *Journal of Chemical Theory and Computation* **2019**, *15* (9), 4721-4734.

12.   Fales, B. S.; Curtis, E. R.; Johnson, K. G.; Lahana, D.; Seritan, S.; Wang, Y.; Weir, H.; Martínez, T. J.; Hohenstein, E. G., Performance of Coupled-Cluster Singles and Doubles on Modern Stream Processing Architectures. *Journal of Chemical Theory and Computation* **2020,** *16* (7), 4021-4028.

13.   Gyevi-Nagy, L.; Kállay, M.; Nagy, P. R., Integral-Direct and Parallel Implementation of the CCSD(T) Method: Algorithmic Developments and Large-Scale Applications. *Journal of Chemical Theory and Computation* **2020,** *16* (1), 366-384.

14.   Datta, D.; Gordon, M. S., A Massively Parallel Implementation of the CCSD(T) Method Using the Resolution-of-the-Identity Approximation and a Hybrid Distributed/Shared Memory Parallelization Model. *Journal of Chemical Theory and Computation* **2021,** *17* (8), 4799-4822.

15. Anisimov, V. M.; Bauer, G. H.; Chadalavada, K.; Olson, R. M.; Glenski, J. W.; Kramer, W. T. C.; Aprà, E.; Kowalski, K., Optimization of the Coupled Cluster Implementation in NWChem on Petascale Parallel Architectures. *Journal of Chemical Theory and Computation* **2014,** *10* (10), 4307-4316.

16. Kaliman, I. A.; Krylov, A. I., New algorithm for tensor contractions on multi-core CPUs, GPUs, and accelerators enables CCSD and EOM-CCSD calculations with over 1000 basis functions on a single compute node. *Journal of Computational Chemistry* **2017,** *38* (11), 842-853.

17. Whitten, J. L., Coulombic potential energy integrals and approximations. *The Journal of Chemical Physics* **1973,** *58* (10), 4496-4501.

18. Almlöf, J.; Faegri Jr, K.; Korsell, K., Principles for a direct SCF approach to LICAO–MOab-initio calculations. *Journal of Computational Chemistry* **1982,** *3* (3), 385-399.

19. Dunlap, B. I.; Connolly, J. W. D.; Sabin, J. R., On the applicability of LCAO-Xα methods to molecules containing transition metal atoms: The nickel atom and nickel hydride. *International Journal of Quantum Chemistry* **1977,** *12* (S11), 81-87.

20. Dunlap, B. I.; Connolly, J. W. D.; Sabin, J. R., On some approximations in applications of Xα theory. *The Journal of Chemical Physics* **1979,** *71* (8), 3396-3402.

21. Feyereisen, M.; Fitzgerald, G.; Komornicki, A., Use of approximate integrals in ab initio theory. An application in MP2 energy calculations. *Chem Phys Lett* **1993,** *208* (5), 359-363.

22. Vahtras, O.; Almlöf, J.; Feyereisen, M. W., Integral approximations for LCAO-SCF calculations. *Chem Phys Lett* **1993,** *213* (5), 514-518.

23. Beebe, N. H. F.; Linderberg, J., Simplifications in the generation and transformation of two-electron integrals in molecular calculations. *International Journal of Quantum Chemistry* **1977,** *12* (4), 683-705.

24. Røeggen, I.; Wisløff-Nilssen, E., On the Beebe-Linderberg two-electron integral approximation. *Chem Phys Lett* **1986,** *132* (2), 154-160.

25. Koch, H.; Sánchez de Merás, A.; Pedersen, T. B., Reduced scaling in electronic structure calculations using Cholesky decompositions. *The Journal of Chemical Physics* **2003,** *118* (21), 9481-9484.

26. Friesner, R. A., Solution of the Hartree–Fock equations by a pseudospectral method: Application to diatomic molecules. *The Journal of Chemical Physics* **1986,** *85* (3), 1462-1468.

27. Friesner, R. A., Solution of the Hartree–Fock equations for polyatomic molecules by a pseudospectral method. *The Journal of Chemical Physics* **1987,** *86* (6), 3522-3531.

28. Langlois, J. M.; Muller, R. P.; Coley, T. R.; Goddard, W. A.; Ringnalda, M. N.; Won, Y.; Friesner, R. A., Pseudospectral generalized valence‐bond calculations: Application to methylene, ethylene, and silylene. *The Journal of Chemical Physics* **1990,** *92* (12), 7488-7497.

29. Ringnalda, M. N.; Belhadj, M.; Friesner, R. A., Pseudospectral Hartree–Fock theory: Applications and algorithmic improvements. *The Journal of Chemical Physics* **1990,** *93* (5), 3397-3407.

30. Friesner, R. A., New Methods For Electronic Structure Calculations on Large Molecules. *Annual Review of Physical Chemistry* **1991,** *42* (1), 341-367.

31. Neese, F.; Wennmohs, F.; Hansen, A.; Becker, U., Efficient, approximate and parallel Hartree–Fock and hybrid DFT calculations. A 'chain-of-spheres' algorithm for the Hartree–Fock exchange. *Chemical Physics* **2009,** *356* (1-3), 98-109.

32. Izsák, R.; Neese, F., An overlap fitted chain of spheres exchange method. *The Journal of chemical physics* **2011,** *135* (14), 144105.

33. Izsák, R.; Neese, F.; Klopper, W., Robust fitting techniques in the chain of spheres approximation to the Fock exchange: The role of the complementary space. *The Journal of chemical physics* **2013,** *139* (9), 094111.

34. Hohenstein, E. G.; Parrish, R. M.; Martinez, T. J., Tensor hypercontraction density fitting. I. Quartic scaling second- and third-order Moller-Plesset perturbation theory. *Journal of Chemical Physics* **2012,** *137* (4), 044103.

35. Parrish, R. M.; Hohenstein, E. G.; Martinez, T. J.; Sherrill, C. D., Tensor hypercontraction. II. Least-squares renormalization. *Journal of Chemical Physics* **2012,** *137* (22).

36. Parrish, R. M.; Hohenstein, E. G.; Schunck, N. F.; Sherrill, C. D.; Martinez, T. J., Exact Tensor Hypercontraction: A Universal Technique for the Resolution of Matrix Elements of Local Finite-Range N-Body Potentials in Many-Body Quantum Problems. *Physical Review Letters* **2013,** *111* (13).

37. Scuseria, G. E.; Ayala, P. Y., Linear scaling coupled cluster and perturbation theories in the atomic orbital basis. *The Journal of chemical physics* **1999,** *111* (18), 8330-8343.

38. Flocke, N.; Bartlett, R. J., A natural linear scaling coupled-cluster method. *The Journal of chemical physics* **2004,** *121* (22), 10935-10944.

39. Pulay, P., Localizability of dynamic electron correlation. *Chem Phys Lett* **1983,** *100* (2), 151-154.

40. Saebo, S.; Pulay, P., The local correlation treatment. II. Implementation and tests. *The Journal of chemical physics* **1988,** *88* (3), 1884-1890.

41. Saebo, S.; Pulay, P., Local configuration interaction: An efficient approach for larger molecules. *Chem Phys Lett* **1985,** *113* (1), 13-18.

42. Saebo, S.; Pulay, P., Local treatment of electron correlation. *Annual Review of Physical Chemistry* **1993,** *44* (1), 213-236.

43. Hetzer, G.; Pulay, P.; Werner, H.-J., Multipole approximation of distant pair energies in local MP2 calculations. *Chem Phys Lett* **1998,** *290* (1-3), 143-149.

44. Hampel, C.; Werner, H. J., Local treatment of electron correlation in coupled cluster theory. *Journal of Chemical Physics* **1996,** *104* (16), 6286-6297.

45. Schutz, M.; Werner, H. J., Local perturbative triples correction (T) with linear cost scaling. *Chem Phys Lett* **2000,** *318* (4-5), 370-378.

46. Schutz, M.; Werner, H. J., Low-order scaling local electron correlation methods. IV. Linear scaling local coupled-cluster (LCCSD). *Journal of Chemical Physics* **2001,** *114* (2), 661-681.

47. Korona, T.; Pfluger, K.; Werner, H. J., The effect of local approximations in coupled-cluster wave functions on dipole moments and static dipole polarisabilities. *Physical Chemistry Chemical Physics* **2004,** *6* (9), 2059-2065.

48. Russ, N. J.; Crawford, T. D., Local correlation in coupled cluster calculations of molecular response properties. *Chem Phys Lett* **2004,** *400* (1-3), 104-111.

49. Crawford, T. D.; King, R. A., Locally correlated equation-of-motion coupled cluster theory for the excited states of large molecules. *Chem Phys Lett* **2002,** *366* (5-6), 611-622.

50. Crawford, T. D., Reduced-Scaling Coupled-Cluster Theory for Response Properties of Large Molecules. In *Recent Progress in Coupled Cluster Methods: Theory and Applications*, Carsky, P.; Paldus, J.; Pittner, J., Eds. 2010; Vol. 11, pp 37-55.

51. Korona, T.; Kats, D.; Schuetz, M.; Adler, T. B.; Liu, Y.; Werner, H.-J., Local Approximations for an Efficient and Accurate Treatment of Electron Correlation and Electron Excitations in Molecules. In *Linear-Scaling Techniques in Computational Chemistry and*

*Physics: Methods and Applications*, Zalesny, R.; Papadopoulos, M. G.; Mezey, P. G.; Leszczynski, J., Eds. 2011; Vol. 13, pp 345-407.

52. Crawford, T. D.; Kumar, A.; Bazante, A. P.; Di Remigio, R., Reduced-scaling coupled cluster response theory: Challenges and opportunities. *Wiley Interdisciplinary Reviews-Computational Molecular Science* **2019,** *9* (4).

53. Li, S.; Shen, J.; Li, W.; Jiang, Y., An efficient implementation of the "cluster-in-molecule" approach for local electron correlation calculations. *The Journal of chemical physics* **2006,** *125* (7), 074109.

54. Li, W.; Piecuch, P., Improved design of orbital domains within the cluster-in-molecule local correlation framework: Single-environment cluster-in-molecule ansatz and its application to local coupled-cluster approach with singles and doubles. *The Journal of Physical Chemistry A* **2010,** *114* (33), 8644-8657.

55. Rolik, Z.; Kállay, M., A general-order local coupled-cluster method based on the cluster-in-molecule approach. *The Journal of chemical physics* **2011,** *135* (10), 104111.

56. Fedorov, D. G.; Kitaura, K., Coupled-cluster theory based upon the fragment molecular-orbital method. *The Journal of Chemical Physics* **2005,** *123* (13), 134103.

57. Azar, R. J.; Head-Gordon, M., An energy decomposition analysis for intermolecular interactions from an absolutely localized molecular orbital reference at the coupled-cluster singles and doubles level. *The Journal of chemical physics* **2012,** *136* (2), 024103.

58. Ghosh, S.; Neese, F.; Izsák, R.; Bistoni, G., Fragment-Based Local Coupled Cluster Embedding Approach for the Quantification and Analysis of Noncovalent Interactions: Exploring the Many-Body Expansion of the Local Coupled Cluster Energy. *Journal of chemical theory and computation* **2021**.

59. Neese, F.; Hansen, A.; Liakos, D. G., Efficient and accurate approximations to the local coupled cluster singles doubles method using a truncated pair natural orbital basis. *Journal of Chemical Physics* **2009,** *131* (6).

60. Neese, F.; Wennmohs, F.; Hansen, A., Efficient and accurate local approximations to coupled-electron pair approaches: An attempt to revive the pair natural orbital method. *Journal of Chemical Physics* **2009,** *130* (11).

61. Hansen, A.; Liakos, D. G.; Neese, F., Efficient and accurate local single reference correlation methods for high-spin open-shell molecules using pair natural orbitals. *Journal of Chemical Physics* **2011,** *135* (21).

62. Riplinger, C.; Sandhoefer, B.; Hansen, A.; Neese, F., Natural triple excitations in local coupled cluster calculations with pair natural orbitals. *Journal of Chemical Physics* **2013,** *139* (13).

63. Ma, Q. L.; Schwilk, M.; Koppl, C.; Werner, H. J., Scalable Electron Correlation Methods. 4. Parallel Explicitly Correlated Local Coupled Cluster with Pair Natural Orbitals (PNO-LCCSD-F12). *Journal of Chemical Theory and Computation* **2017,** *13* (10), 4871-4896.

64. Schwilk, M.; Ma, Q. L.; Koppl, C.; Werner, H. J., Scalable Electron Correlation Methods. 3. Efficient and Accurate Parallel Local Coupled Cluster with Pair Natural Orbitals (PNO-LCCSD). *Journal of Chemical Theory and Computation* **2017,** *13* (8), 3650-3675.

65. Ma, Q. L.; Werner, H. J., Explicitly correlated local coupled-cluster methods using pair natural orbitals. *Wiley Interdisciplinary Reviews-Computational Molecular Science* **2018,** *8* (6).

66. Ma, Q. L.; Werner, H. J., Scalable Electron Correlation Methods. 5. Parallel Perturbative Triples Correction for Explicitly Correlated Local Coupled Cluster with Pair Natural Orbitals. *Journal of Chemical Theory and Computation* **2018,** *14* (1), 198-215.

67. Werner, H. J., Communication: Multipole approximations of distant pair energies in local correlation methods with pair natural orbitals. *Journal of Chemical Physics* **2016,** *145* (20).

68. Meyer, W., Ionization energies of water from PNO‐CI calculations. *International Journal of Quantum Chemistry* **1971,** *5* (S5), 341-348.

69. Meyer, W., PNO–CI studies of electron correlation effects. I. Configuration expansion by means of nonorthogonal orbitals, and application to the ground state and ionized states of methane. *The Journal of Chemical Physics* **1973,** *58* (3), 1017-1035.

70. Meyer, W.; Rosmus, P., PNO–CI and CEPA studies of electron correlation effects. III. Spectroscopic constants and dipole moment functions for the ground states of the first‐row and second‐row diatomic hydrides. *The Journal of Chemical Physics* **1975,** *63* (6), 2356-2375.

71. Werner, H.-J.; Meyer, W., PNO-CI and PNO-CEPA studies of electron correlation effects: V. Static dipole polarizabilities of small molecules. *Molecular Physics* **1976,** *31* (3), 855-872.

72. Riplinger, C.; Neese, F., An efficient and near linear scaling pair natural orbital based local coupled cluster method. *Journal of Chemical Physics* **2013,** *138* (3).

73. Sparta, M.; Neese, F., Chemical applications carried out by local pair natural orbital based coupled-cluster methods. *Chemical Society Reviews* **2014,** *43* (14), 5032-5041.

74. Helmich, B.; Haettig, C., Local pair natural orbitals for excited states. *Journal of Chemical Physics* **2011,** *135* (21).

75. Dutta, A. K.; Neese, F.; Izsak, R., Towards a pair natural orbital coupled cluster method for excited states. *Journal of Chemical Physics* **2016,** *145* (3).

76. Frank, M. S.; Haettig, C., A pair natural orbital based implementation of CCSD excitation energies within the framework of linear response theory. *Journal of Chemical Physics* **2018,** *148* (13).

77. Peng, C.; Clement, M. C.; Valeev, E. F., State-Averaged Pair Natural Orbitals for Excited States: A Route toward Efficient Equation of Motion Coupled-Cluster. *Journal of Chemical Theory and Computation* **2018,** *14* (11), 5597-5607.

78. Rendell, A. P.; Lee, T. J., Coupled‐cluster theory employing approximate integrals: An approach to avoid the input/output and storage bottlenecks. *The Journal of Chemical Physics* **1994,** *101* (1), 400-408.

79. DePrince, A. E.; Sherrill, C. D., Accuracy and Efficiency of Coupled-Cluster Theory Using Density Fitting/Cholesky Decomposition, Frozen Natural Orbitals, and a t(1)-Transformed Hamiltonian. *Journal of Chemical Theory and Computation* **2013,** *9* (6), 2687-2696.

80. Epifanovsky, E.; Zuev, D.; Feng, X.; Khistyaev, K.; Shao, Y.; Krylov, A. I., General implementation of the resolution-of-the-identity and Cholesky representations of electron repulsion integrals within coupled-cluster and equation-of-motion methods: Theory and benchmarks. *The Journal of Chemical Physics* **2013,** *139* (13), 134105.

81. Parrish, R. M.; Sherrill, C. D.; Hohenstein, E. G.; Kokkila, S. I. L.; Martinez, T. J., Communication: Acceleration of coupled cluster singles and doubles via orbital-weighted least-squares tensor hypercontraction. *Journal of Chemical Physics* **2014,** *140* (18).

82. Dutta, A. K.; Neese, F.; Izsak, R., Speeding up equation of motion coupled cluster theory with the chain of spheres approximation. *Journal of Chemical Physics* **2016,** *144* (3).

83. Dutta, A. K.; Neese, F.; Izsak, R., Accelerating the coupled-cluster singles and doubles method using the chain-of-sphere approximation. *Molecular Physics* **2018,** *116* (11), 1428-1434.

84. Pierce, K.; Rishi, V.; Valeev, E. F., Robust Approximation of Tensor Networks: Application to Grid-Free Tensor Factorization of the Coulomb Interaction. *Journal of Chemical Theory and Computation* **2021,** *17* (4), 2217-2230.

85. Hummel, F.; Tsatsoulis, T.; Grüneis, A., Low rank factorization of the Coulomb integrals for periodic coupled cluster theory. *The Journal of chemical physics* **2017,** *146* (12), 124105.

86. Sun, Q.; Berkelbach, T. C.; McClain, J. D.; Chan, G. K.-L., Gaussian and plane-wave mixed density fitting for periodic systems. *The Journal of chemical physics* **2017,** *147* (16), 164119.

87. Ye, H.-Z.; Berkelbach, T. C., Fast periodic Gaussian density fitting by range separation. *The Journal of Chemical Physics* **2021,** *154* (13), 131104.

88. Hohenstein, E. G.; Parrish, R. M.; Sherrill, C. D.; Martinez, T. J., Communication: Tensor hypercontraction. III. Least-squares tensor hypercontraction for the determination of correlated wavefunctions. *Journal of Chemical Physics* **2012,** *137* (22).

89. Carroll, J. D.; Chang, J.-J., Analysis of individual differences in multidimensional scaling via an N-way generalization of "Eckart-Young" decomposition. *Psychometrika* **1970,** *35* (3), 283-319.

90. Harshman, R. A., Foundations of the PARAFAC procedure: Models and conditions for an" explanatory" multimodal factor analysis. *UCLA Working Papers in Phonetics* **1970,** *16*, 1.

91. Benedikt, U.; Bohm, K. H.; Auer, A. A., Tensor decomposition in post-Hartree-Fock methods. II. CCD implementation. *Journal of Chemical Physics* **2013,** *139* (22).

92. Schutski, R.; Zhao, J. M.; Henderson, T. M.; Scuseria, G. E., Tensor-structured coupled cluster theory. *Journal of Chemical Physics* **2017,** *147* (18).

93. Hohenstein, E. G.; Zhao, Y.; Parrish, R. M.; Martinez, T. J., Rank reduced coupled cluster theory. II. Equation-of-motion coupled-cluster singles and doubles. *J Chem Phys* **2019,** *151* (16), 164121.

94. Parrish, R. M.; Zhao, Y.; Hohenstein, E. G.; Martinez, T. J., Rank-Reduced Coupled-Cluster Theory I. Ground State Energies and Wavefunctions. *Journal of Chemical Physics* **2019,** *150*, 164118.

95. Stanton, J. F.; Bartlett, R. J., The Equation of Motion Coupled-Cluster Method. A Systematic Biorthogonal Approach to Molecular-Excitation Energies, Transition-Probabilities, and Excited-State Properties. *Journal of Chemical Physics* **1993,** *98* (9), 7029-7039.

96. Hohenstein, E. G.; Martinez, T. J., GPU Acceleration of Rank-Reduced Coupled-Cluster Singles and Doubles. *J Chem Phys*, in press.

97. Kinoshita, T.; Hino, O.; Bartlett, R. J., Singular value decomposition approach for the approximate coupled-cluster method. *Journal of Chemical Physics* **2003,** *119* (15), 7756-7762.

98. Hino, O.; Kinoshita, T.; Bartlett, R. J., Singular value decomposition applied to the compression of T-3 amplitude for the coupled cluster method. *Journal of Chemical Physics* **2004,** *121* (3), 1206-1213.

99. Tucker, L. R., Some mathematical notes on three-mode factor analysis. *Psychometrika* **1966,** *31* (3), 279-311.

100. Lesiuk, M., Efficient singular‐value decomposition of the coupled‐cluster triple excitation amplitudes. *Journal of computational chemistry* **2019,** *40* (12), 1319-1332.

101. Lesiuk, M., Implementation of the Coupled-Cluster Method with Single, Double, and Triple Excitations using Tensor Decompositions. *Journal of chemical theory and computation* **2019,** *16* (1), 453-467.

102. Kolda, T. G.; Bader, B. W., Tensor decompositions and applications. *SIAM review* **2009,** *51* (3), 455-500.

103. Hu, W.; Lin, L.; Yang, C., Interpolative Separable Density Fitting Decomposition for Accelerating Hybrid Density Functional Calculations with Applications to Defects in Silicon. *Journal of Chemical Theory and Computation* **2017,** *13* (11), 5420-5431.

104. Lu, J.; Ying, L., Compression of the electron repulsion integral tensor in tensor hypercontraction format with cubic scaling cost. *Journal of Computational Physics* **2015,** *302*, 329-335.

105. Lee, J.; Lin, L.; Head-Gordon, M., Systematically Improvable Tensor Hypercontraction: Interpolative Separable Density-Fitting for Molecules Applied to Exact Exchange, Second- and Third-Order Moller-Plesset Perturbation Theory. *Journal of Chemical Theory and Computation* **2020,** *16* (1), 243-263.

106. Pierce, K.; Rishi, V.; Valeev, E. F., Robust approximation of tensor networks: application to grid-free tensor factorization of the Coulomb interaction. *arXiv preprint arXiv:2012.13002* **2020**.

107. Koch, H.; Christiansen, O.; Kobayashi, R.; Jorgensen, P.; Helgaker, T., A Direct Atomic Orbital Driven Implementation of the Coupled-Cluster Singles and Doubles (CCSD) Model. *Chem Phys Lett* **1994,** *228* (1-3), 233-238.

108. Ufimtsev, I. S.; Martínez, T. J., Quantum Chemistry on Graphical Processing Units. 1. Strategies for Two-Electron Integral Evaluation. *Journal of Chemical Theory and Computation* **2008,** *4* (2), 222-231.

109. Ufimtsev, I. S.; Martinez, T. J., Quantum Chemistry on Graphical Processing Units. 2. Direct Self-Consistent-Field Implementation. *Journal of Chemical Theory and Computation* **2009,** *5* (4), 1004-1015.

110. Ufimtsev, I. S.; Martinez, T. J., Quantum Chemistry on Graphical Processing Units. 3. Analytical Energy Gradients, Geometry Optimization, and First Principles Molecular Dynamics. *Journal of Chemical Theory and Computation* **2009,** *5* (10), 2619-2628.

111. Titov, A. V.; Ufimtsev, I. S.; Luehr, N.; Martinez, T. J., Generating Efficient Quantum Chemistry Codes for Novel Architectures. *Journal of Chemical Theory and Computation* **2013,** *9* (1), 213-221.

112. Lesiuk, M., Quintic-scaling rank-reduced coupled cluster theory with single and double excitations. *arXiv preprint arXiv:2109.08583* **2021**.

113. DiStasio, R. A.; Jung, Y.; Head-Gordon, M., A Resolution-Of-The-Identity Implementation of the Local Triatomics-In-Molecules Model for Second-Order Møller−Plesset Perturbation Theory with Application to Alanine Tetrapeptide Conformational Energies. *Journal of Chemical Theory and Computation* **2005,** *1* (5), 862-876.